\documentclass[twocolumn,preprintnumbers,amsmath,amssymb,prl]{revtex4}

\usepackage{graphicx}
\usepackage{dcolumn}
\usepackage{bm}

\begin{document}

\title{Direct observation of ultrafast many-body electron dynamics\\ in an ultracold Rydberg gas}

\author{Nobuyuki Takei$^{1,2,^{*}}$, Christian Sommer$^{1,2,^{*}}$, Claudiu Genes$^{3}$, Guido Pupillo$^{4}$,\\
Haruka Goto$^{1}$, Kuniaki Koyasu$^{1,2}$, Hisashi Chiba$^{1,5}$,
Matthias Weidem\"uller$^{6,7,8}$ and Kenji Ohmori$^{1,2}$}

\affiliation{%
$^{1}$Department of Photo-Molecular Science, Institute for Molecular Science, National Institutes of Natural Sciences, Myodaiji, Okazaki 444-8585, Japan\\
$^{2}$SOKENDAI (The Graduate University for Advanced Studies), Myodaiji, Okazaki 444-8585, Japan\\
$^{3}$Institut f\"ur Theoretische Physik, Universit\"at Innsbruck, Technikerstrasse 25, A-6020 Innsbruck, Austria\\
$^{4}$IPCMS (UMR 7504) and ISIS (UMR 7006), University of Strasbourg and CNRS, 67000 Strasbourg, France\\
$^{5}$Faculty of Engineering, Iwate University, 4-3-5 Ueda, Morioka 020-8551, Japan\\
$^{6}$Physikalisches Institut, Universit\"at Heidelberg, Im Neuenheimer Feld 226, 69120 Heidelberg, Germany\\
$^{7}$Hefei National Laboratory for Physical Sciences at the Microscale and Department of Modern Physics, University of Science and Technology of China, Hefei, Anhui 230026, China\\
$^{8}$CAS Center for Excellence and Synergetic Innovation Center in Quantum Information and Quantum Physics, University of Science and Technology of China, Hefei, Anhui 230026, China\\
$^{*}$These authors contributed equally to this work.
}%

\date{\today}

\begin{abstract}

Many-body correlations govern a variety of important quantum phenomena such as the emergence of superconductivity and magnetism.
Understanding quantum many-body systems is thus one of the central goals of modern sciences.
Here we demonstrate an experimental approach towards this goal by utilizing an ultracold Rydberg gas generated with a broadband picosecond laser pulse. We follow the ultrafast evolution of its electronic coherence by time-domain Ramsey interferometry with attosecond precision. The observed electronic coherence shows an ultrafast oscillation with a period of 1\,femtosecond, whose phase shift on the attosecond timescale is consistent with many-body correlations among Rydberg atoms beyond mean-field approximations. This coherent and ultrafast many-body dynamics is actively controlled by tuning the orbital size and population of the Rydberg state, as well as the mean atomic distance.
Our approach will offer a versatile platform to observe and manipulate non-equilibrium dynamics of quantum many-body systems on the ultrafast timescale.

\end{abstract}

\maketitle

\section{Introduction}

Atomic, molecular and optical physics with advanced laser technologies has recently emerged as a new platform to study and possibly simulate quantum many-body systems \cite{Jaksch1998,Greiner2002,Bloch2008}.
One of its latest developments is the study of long-range interactions among ultracold particles, in which dipolar quantum gases~\cite{Lahaye2009,Pupillo2012}, ion crystals~\cite{Monroe2014,Blatt2014,Bohnet2016}, polar molecules~\cite{Yan2013,Hazzard2014a,Takekoshi2014} and Rydberg atoms~\cite{Saffman2010,Pillet2010,Pfau2014mol,Bloch2016} allow for revealing the effects of many-body correlations. 
Rydberg atoms distinguish themselves by their large dipole moments and tunability of the strength and nature of dipolar interactions~\cite{Saffman2010,Pillet2010}.

\begin{figure*}
	\begin{center}
		\includegraphics[scale=0.8]{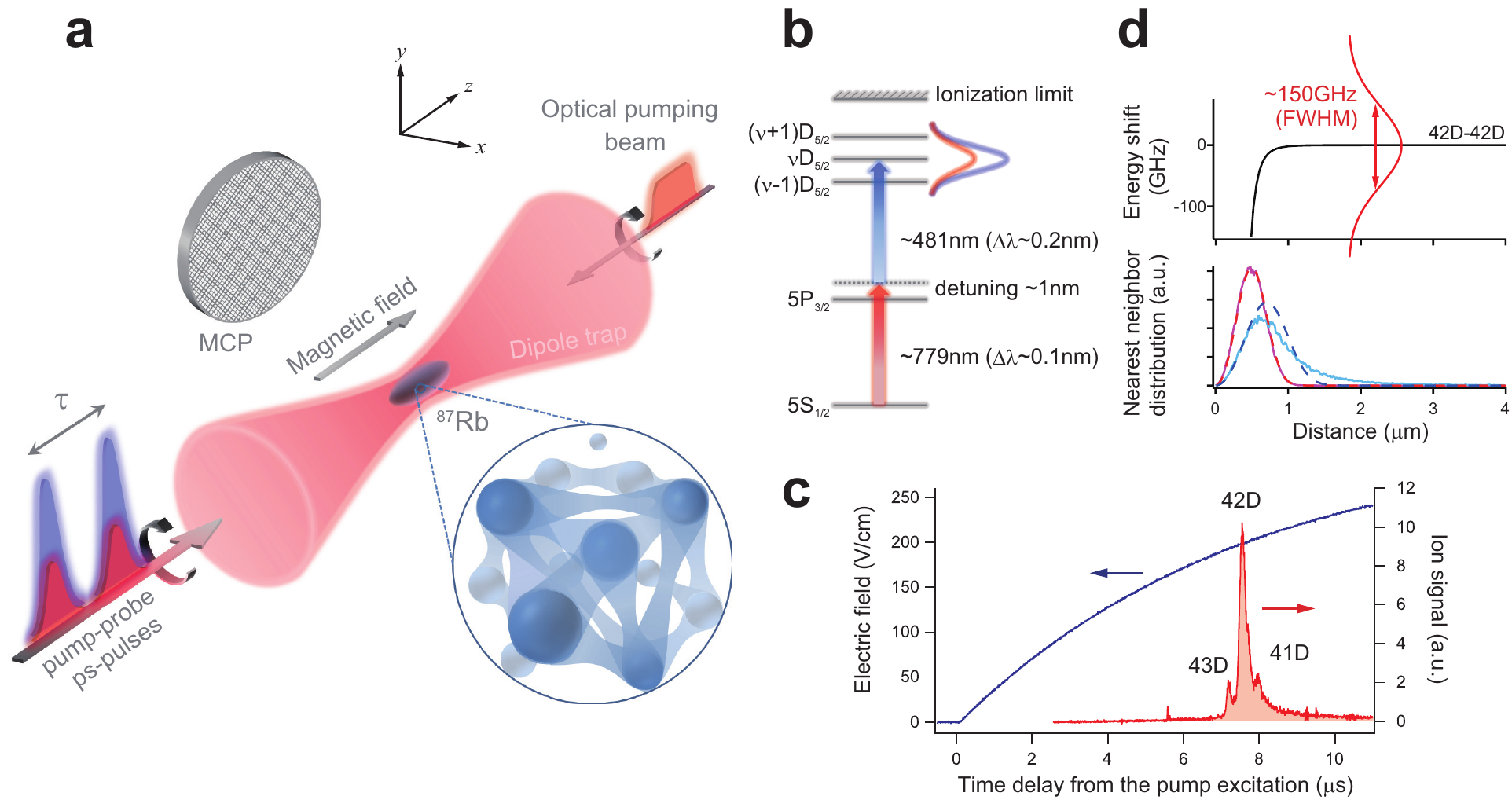}
		\caption{{\bf Schematic diagram of the experimental setup and Rydberg excitation.}
(\textbf{a}) Sketch of the experimental setup.  The dipole-trap laser is turned off 2\,$\mathrm{\mu s}$ before the irradiation of the picosecond pulses, to avoid $2$+$1$ multiphoton ionization induced by a combination of the picosecond pulses and the trapping laser beam.
(\textbf{b}) Two-photon pump (probe) excitation of the Rb atom to its Rydberg states.
(\textbf{c}) The Rydberg states can be resolved by field ionization with a slowly ramped electric field (see Methods section `Rydberg excitation and detection'). Here, the Rydberg population and estimated peak atom density were $1.2\pm0.1\,\%$ and $\sim4\times 10^{10}\,\mathrm{cm}^{-3}$, respectively (see Methods sections `Rydberg excitation and detection' and `Estimation of the atom density'  for these population and density estimations). The small peak around $5.6\,\mu$s could be assigned to free ions generated by population redistribution~\cite{Tanner2008} and/or direct multiphoton ionization.
(\textbf{d}) Sketch of the two-body interaction and pulse excitation accompanied by a plot of the nearest-neighbor distribution of Rydberg atoms for the peak atom density of $n = 1.3\times 10^{12}\,\mathrm{cm}^{-3}$ (pink solid and red dashed traces) and the averaged density over the whole atoms (light-blue solid and dark-blue dashed traces) estimated for the present ensemble of the Rb atoms (see Methods section `Estimation of the atom density'). Here, the pink and light-blue solid traces are obtained by a Monte-Carlo simulation, whereas the red and dark-blue dashed traces show analytical results for homogeneous distributions. The difference between the light-blue and dark-blue average density traces results from the difference between their Gaussian and homogeneous density distributions, respectively, over the whole ensemble. At the peak atom density, the average nearest-neighbour distance is given by $0.5\,\mu$m, whereas for the averaged density we obtain $0.87\,\mu$m.}
		\label{fig01}
	\end{center}
\end{figure*}

The active electron in a Rydberg atom moves in a macroscopic orbital whose size could range from sub-micrometre to several tens of micrometres~\cite{Gallagher1994}. The resulting large dipole moments yield a strong interaction $U(r)$ of either van der Waals or resonant dipolar character between a pair of Rydberg atoms separated by $r$. This interaction, which can be tuned in various ways~\cite{Saffman2010}, features the emergence of atomic many-body correlations in an ultracold gas. As a prominent example, these interactions shift the energy levels of atoms encompassing a given Rydberg atom, so that additional Rydberg excitations are suppressed for $U(r)$ larger than the excitation line width. This effect, referred to as Rydberg blockade~\cite{Gould2004,Saffman2010,Pillet2010}, results in spatial or temporal correlations among the atoms, as recently demonstrated in various settings~\cite{Bloch2016,Schauss2012,Bloch2015,Raithel2011,Nipper2012a,Nipper2012b,Bettelli2013,Browaeys2014,Barredo2014,Browaeys2015}. Its applications include, for example, the realization of universal atomic and photonic logic gates for quantum information processing~\cite{Saffman2010,Jaksch2000,Lukin2001,Dudin2012,Firstenberg2013,Rempe2014}.
The Rydberg blockade has also turned out to be an outstanding new resource to investigate many-body problems~\cite{Weimer2010}.
The blockade condition determines the smallest distance between neighbouring Rydberg atoms, which is typically of the order of several micrometres.
However, Rydberg excitations can also be induced at smaller interatomic distances by, for example, detuning the excitation-laser frequency from an atomic resonance~\cite{Haroche1981,Morsch2012,Morsch2014,Amthor2007,Schempp2014,Urvoy2015}. In this regime, the character of the correlations changes, resulting in the facilitated formation of aggregates consisting of large number of Rydberg atoms~\cite{Amthor2007,Morsch2012,Schempp2014,Morsch2014,Urvoy2015}.

A complementary approach to the correlations induced by the interparticle interactions consists in studying the temporal evolution of electronic coherences of the Rydberg atoms.
Using a broadband picosecond laser pulse,  Rydberg excitations can be induced over a wide range of interatomic distances from $<1\,\mu$m to the isolated atom limit.
The number of Rydberg excitations per unit volume can thus be larger than the ones in the blockade regime by two orders of magnitude.
Although picosecond and femtosecond laser pulses have been previously used to observe electronic wave packets in isolated Rydberg atoms~\cite{Alber1991,Bucksbaum1999} and the dephasing due to two-body interactions~\cite{Zhou2014}, here we exploit them to explore how coherent dynamics evolves in a many-body regime.
The strong interactions in our ultracold Rydberg gas induce an electronic dephasing on the picosecond timescale, which is directly observed in a time-domain Ramsey interferogram oscillating with a period of $\sim1$\,fs and phase-shifted on the attosecond timescale~\cite{Ohmori2009} by the Rydberg interactions. We measure this minute phase shift and the dephasing directly, and compare them with theoretical simulations based on nearest-neighbour interactions, a mean-field model and many-body correlations to reveal effects indicating atomic correlations beyond a mean-field description.
The two-body interaction energy at around $1\,\mu$m distance, which can be easily accessed in our approach, exceeds the average kinetic energy by many orders of magnitude.
This regime compares favorably with previous experiments in thermal vapour cell experiments with nanosecond laser pulses~\cite{Urvoy2015,Baluktsian2013}, where the nearest-neighbour distance is also $<1\,\mu$m. In thermal cells, the Rydberg interaction energy is comparable to the average kinetic energy, so that the measurement of coherent evolution of the interaction dynamics is strongly affected by the thermal atomic motion, as will be discussed later quantitatively.

\section{Results}
\subsection{Experimental set-up}

Figure~\ref{fig01}a shows the schematics of our experimental setup. A cold ensemble of $^{87}$Rb atoms is prepared in an optical dipole trap with temperature and highest peak atom-density estimated to be $\sim$70\,$\mathrm{\mu K}$ and $\sim$1.3$\times10^{12}$\,cm$^{-3}$, respectively (see Methods sections `Estimation of the temperature' and `Estimation of the atom density' for these temperature and density estimations).
On the timescales of interest, atomic motion can be ignored in the frozen gas regime~\cite{Anderson1998,Mourachko1998}.
The atoms are optically pumped to the hyperfine state F=2, m$_{\mathrm{F}}$=+2 of the ground state 5S$_{1/2}$ and excited to Rydberg states via a two-photon transition using broadband picosecond laser pulses with their centre wavelengths tuned to $\sim779$ and $\sim481\,$nm (Fig.~\ref{fig01}b), hereafter referred to as the infrared and blue pulses, respectively.
The dipole-trap laser is turned off $2\,\mu\mathrm{s}$ before the irradiation of the infrared and blue pulses to avoid 2+1 multiphoton ionization induced by a combination of the infrared and blue pulses with the trapping laser light.
The infrared and blue pulses, and the optical pumping beam are circularly polarized in the same direction with respect to the magnetic field, suppressing excitations to the S Rydberg states, so that the state $\nu$D$_{5/2}$, m$_{\mathrm{J}}$=+5/2 is mostly populated, where $\nu$ is a principal quantum number.
The population of the Rydberg state is measured by field ionization~\cite{Gallagher1994}. Details of the field ionization are described in Methods section `Rydberg excitation and detection'.
The maximum population of the $\nu$D states in the present experiment is not $>5\,\%$, to suppress photoionization by the picosecond laser pulses (see Supplementary Note 1 for the effects of ions). More details on the atom preparation, the estimation of the atom density and the temperature, as well as the Rydberg excitation are described in Methods and Supplementary Note 2.

\begin{figure*}
	\begin{center}
		\includegraphics[scale=0.7]{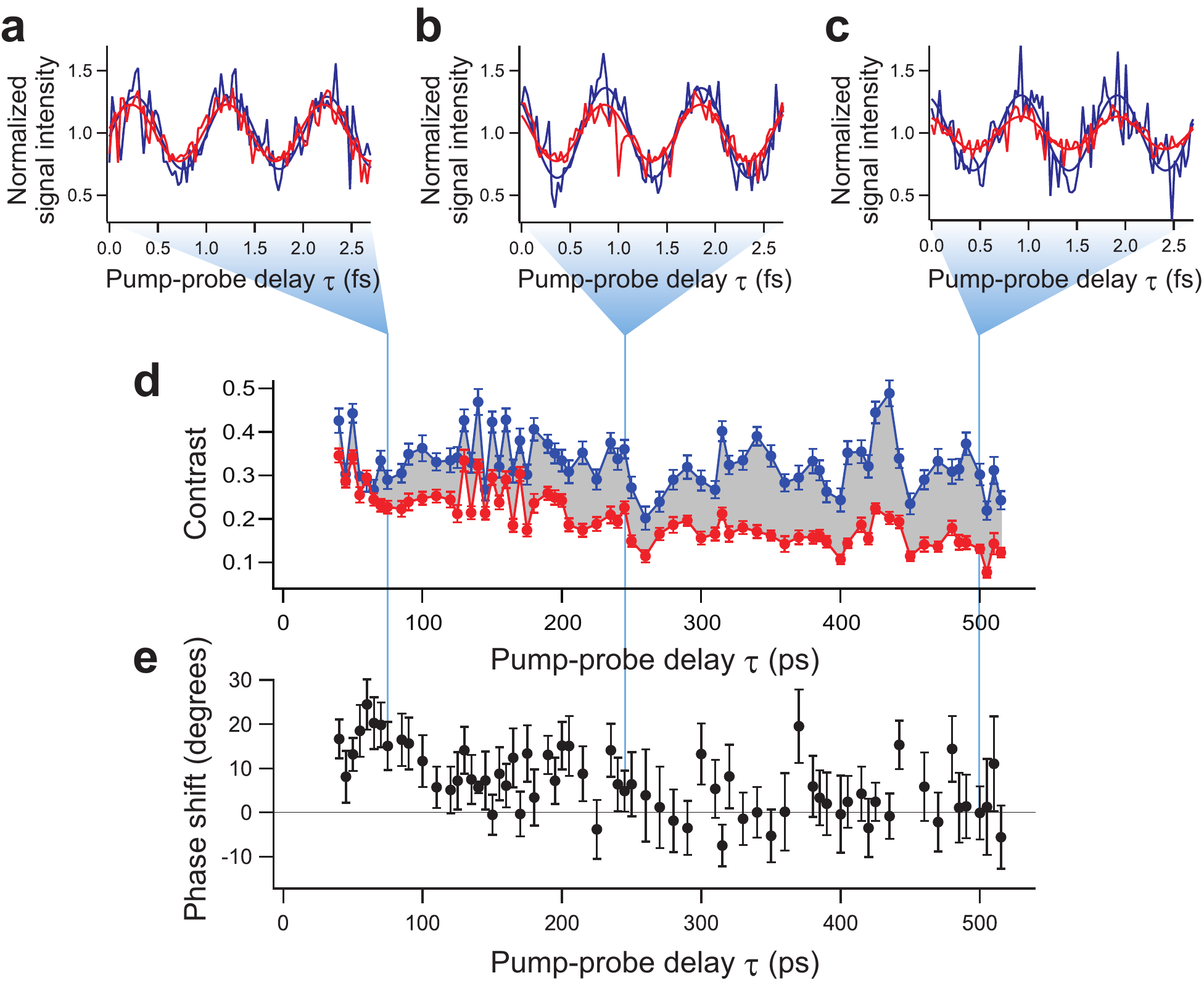}
		\caption{{\bf Ramsey oscillation for the 42D state.}
(\textbf{a} -- \textbf{c}) The field ionization signals for the higher-density (red traces) and the lower-density (blue traces) ensembles are plotted as functions of the pump-probe delay $\tau$ scanned over a range of $\sim3$\,fs around $75$, $245$ and $500$\,ps, showing clear oscillations. The signal intensities are normalized by the mean value of the sinusoidal function fitted to each oscillation. The origin of the pump-probe delay $\tau = 0$ is arbitrary and is taken to be the left edge of each figure.
(\textbf{d}) The contrasts of these oscillations are plotted as functions of $\tau$. 
The oscillatory structures as functions of $\tau$ could be partly attributable to the recurrence motion of a wave packet composed of the $42$D and its neighbouring Rydberg states (see the field-ionization spectrum shown in Fig.~1c). The shortest period of this recurrence motion is evaluated from their level spacing to be $\sim10$\,ps, which is not resolved in these plots. More details on the oscillatory structures are described in Supplementary Note 4. The contrast decays clearly as a function of $\tau$ for the higher-density ensemble, whereas for the lower-density one a decay is not visible.
(\textbf{e}) The phase shift of the higher-density ensemble from the lower-density one is plotted as a function of $\tau$. The oscillatory structure as a function of $\tau$ is partly attributable to the wave-packet motion. The error bars represent the s.d.}
		\label{fig02}
	\end{center}
\end{figure*}

The bandwidth of our Rydberg excitation is determined from a field-ionization spectrum exemplified in Fig.~\ref{fig01}c (see Supplementary Note 3), in which the excitation is tuned to the $42$D$_{5/2}$ state, which is the main target state of the current experiment.
The field-ionization voltage is ramped up slowly enough on the $5\,\mu$s timescale to resolve neighbouring Rydberg levels in such field-ionization spectra.
The bandwidth of the excitation is $\sim 150$\,GHz (full width at half maximum, FWHM) and is much larger than those of the continuous-wave, nanosecond, and sub-nanosecond pulsed lasers employed in previous ultracold Rydberg experiments~\cite{Saffman2010,Pillet2010,Zhou2014,Kutteruf2012,Tanner2008}.
The field-ionization spectrum indicates that the bandwidth of the Rydberg excitation is larger than the energy separation of neighbouring Rydberg states and is accordingly wide enough to remove the Rydberg blockade.
As schematically shown in the upper panel of Fig.~\ref{fig01}d, our picosecond laser pulses can excite a pair of Rb atoms simultaneously to the Rydberg states $\nu=42$ even at interatomic distances shorter than $1\,\mu$m.

\subsection{Observation of many-body electron dynamics}

The interaction among the Rydberg atoms is observed by time-domain Ramsey interferometry with a pair of the Rydberg excitations, hereafter referred to as the `pump' and `probe' excitations, whose delay was stabilized on the attosecond timescale with our homemade optical interferometer~\cite{Katsuki2013} (see Methods section `Time-domain Ramsey interferometry'). By scanning the delay time $\tau$ between the pump and probe excitations on the attosecond timescale, we measure the Ramsey oscillation of the population integrated over all Rydberg states, which remains after the probe excitation, by the field ionization.
The field-ionization voltage in these Ramsey measurements is ramped up rapidly enough on the $100$\,ns timescale to avoid subsequent decay processes of the Rydberg states. For the excitation tuned to the $42$D$_{5/2}$ state as shown in Fig. \ref{fig01}c, a single Rydberg state $|\nu\rangle$ with $\nu=42$ is predominantly populated. 
The Rydberg population thus measured oscillates as a function of $\tau$ with a frequency close to the transition frequency between the 5S and Rydberg states~\cite{Ohmori2009}. This is in contrast to the standard Ramsey interferometry, in which the signal oscillates with a frequency close to the detuning frequency of the excitation laser from the atomic transition~\cite{Nipper2012a,Nipper2012b}.
That is, in the absence of interactions, the population in the Rydberg state $P_\nu(\tau)$ is given by
\begin{eqnarray}
\label{eq:main1}P_\nu(\tau) \propto 1 + \cos(E_{\nu}\tau / \hbar)
\end{eqnarray}
and oscillates with the frequency $E_{\nu}/h$, where $E_{\nu}$ is the energy of the Rydberg state $|\nu\rangle$ measured from the ground state 5S (see refs 42, 49 and Supplementary Note 2), and $\hbar$ is the Planck constant $h$ divided by $2\pi$.
This oscillation is identical to the temporal oscillation of the Rydberg state $|\nu\rangle$, except that the real-time $t$ is replaced by the pump-probe delay $\tau$. Therefore, the Ramsey oscillation for $\nu=42$ corresponds  essentially to the temporal oscillation of the Rydberg wave function $|\nu=42\rangle$ and to the recurrence motion of an electronic wave packet, which is composed of the $5$S and Rydberg state $\nu=42$ superposed coherently by the excitation pulses.
Here we investigate how the Rydberg interactions affect this coherent electron dynamics.

In a simplified mean-field approach, the  Rydberg interactions change only the energy of the Rydberg state $|\nu\rangle$ and  accordingly the oscillation period of $P_\nu(\tau)$. This results in a phase-shift of the Ramsey oscillation  accumulated in the delay time $\tau$. In addition, as the atoms are randomly distributed in the ensemble, their energy levels $E_{\nu}$ are shifted randomly by the interactions, making the periods of their Ramsey oscillations different from each other. Therefore, the measured signal is the superposition of many oscillations with different periods, so that its contrast is expected to decay as a function of  $\tau$ due to Rydberg interactions.

Figure~\ref{fig02}a-c show examples of the Ramsey oscillations for the $42$D$_{5/2}$ state with a population of $3.3\pm0.1\,\%$.
As mentioned above, these oscillations correspond to the ultrafast recurrence motion of the electronic wave packet with a period of $\sim1$\,fs.
For each oscillation, we measured the field-ionization signals of two atomic ensembles with different peak densities estimated to be $\sim1.3\times10^{12}$ and $\sim4\times10^{10}$\,cm$^{-3}$ alternately to suppress systematic uncertainties, scanning $\tau$ in steps of $\sim30$\,as
(see Methods section `Estimation of the atom density' for these density estimations).
We obtained the contrasts and phases of the measured oscillations by sinusoidal fitting, as shown in Fig. \ref{fig02}a-c (see Methods section `Time-domain Ramsey interferometry').
Figure~\ref{fig02}d shows that the contrast is approximately constant for $\tau$ up to $\sim500\,$ps for the lower-density ensemble (blue-circle data points), indicating that the effects of the interactions are negligibly small.
This result is consistent with the interaction strength estimated from the present atom density and the two-atom potential curve presented in Supplementary Fig.~1.
Hereafter, we take these contrasts and phases measured in the lower-density ensemble as references to be compared with those measured in the higher-density ensemble.
The oscillatory structures  on the $\sim10$\,ps timescale at $\tau \sim 130 -170$\,ps seen in Fig.~2d could be partly due to the recurrence motion of a wave-packet composed of the Rydberg state $\nu=42$ and the traces of its neighbouring states seen in Fig.~1c (see Supplementary Note 4 for details).

Figure~\ref{fig02}d shows that the contrast decays as a function of $\tau$ for the higher-density ensemble (red-circle data points). The phase shift of the higher-density ensemble from the lower-density one also changes as a function of $\tau$ as seen in Fig.~\ref{fig02}e. The offset of this phase shift at $\tau=0$ is essentially due to the difference between AC-Stark shifts of the atomic levels in the higher- and lower-density ensembles.
Slight differences between the sizes, shapes and positions of the higher- and lower-density ensembles could lead to their different AC-Stark shifts (see Supplementary Note 2 for more details on the origin of this zero-delay offset of the phase shift).

\begin{figure*}
	\begin{center}
		\includegraphics[scale=0.8]{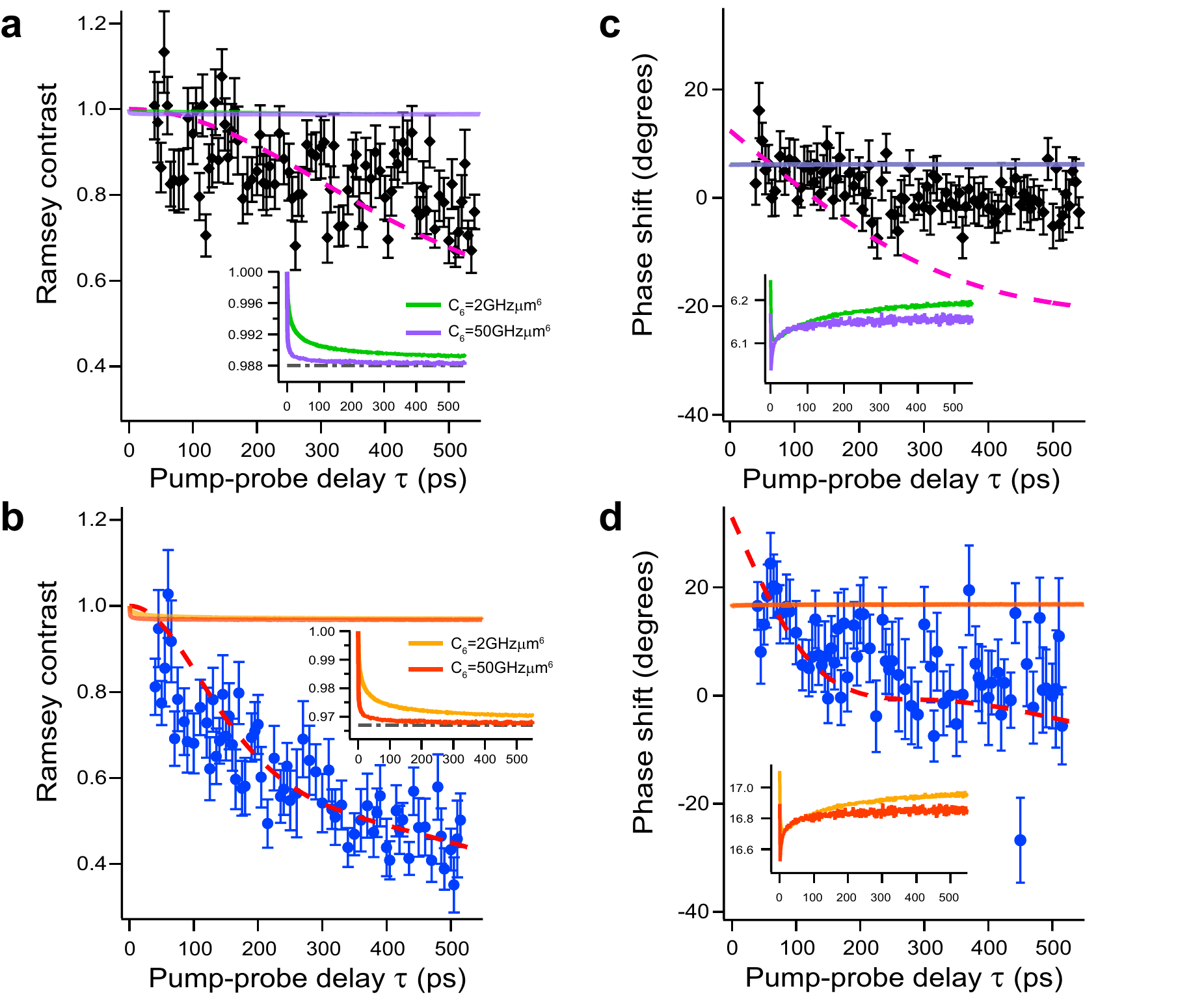}
		\caption{{\bf Measured Ramsey contrasts and phase shifts compared with those expected for nearest-neighbour interactions and mean-field approximations.}
The black diamond-shaped and blue circle data points show the measured Ramsey contrasts (\textbf{a} and \textbf{b}) and the phase shifts (\textbf{c} and \textbf{d}) for the population $p_\mathrm{e}\sim1.2\,\%$ (\textbf{a} and \textbf{c}) and $\sim3.3\,\%$ (\textbf{b} and \textbf{d}) of the $42$D$_{5/2}$ state, respectively. In \textbf{a} and \textbf{b}, the measured Ramsey contrasts are compared with the simulated ones for nearest-neighbour interactions with a $C_6$ coefficient of $2\,\mathrm{GHz}\,\mu\mathrm{m}^{6}$ (green and yellow solid lines) and $50\,\mathrm{GHz}\,\mu\mathrm{m}^{6}$ (purple and orange solid lines), respectively. Results of the mean-field simulations are presented by magenta and red dashed lines in \textbf{a} and \textbf{b}, respectively. Similarly, the measured and simulated phase shifts are compared in \textbf{c} and \textbf{d}. The zero-delay offset of the simulated phase shift is arbitrary and adjusted so that the average of the first ten data points is equal to the calculated phase shift averaged over the delay window for those ten data points. In the insets in \textbf{a} and \textbf{b}, the Ramsey contrasts simulated for the nearest-neighbour interactions (solid lines) are vertically magnified to show their convergence as the $C_6$ value is increased. Grey dot-and-dash lines show their population-dependent thresholds given by $1-p_\mathrm{e}$. In the insets in \textbf{c} and \textbf{d}, the phase shifts simulated for the nearest-neighbour interactions (solid lines) are magnified vertically. The peak atom density is set to $\sim1.3\times10^{12}\,\mathrm{cm}^{-3}$ in these simulations. The error bars represent the s.d.}
		\label{fig03}
	\end{center}
\end{figure*}

Figure~\ref{fig03} shows the results of measurements of the contrast ratio between the two different densities given above, hereafter referred to as `Ramsey contrast',  and the phase shift for the 42D$_{5/2}$ state. These results are plotted as functions of $\tau$ for the two different Rydberg populations $p_\mathrm{e}$ of 1.2$\pm$0.1$\%$ and 3.3$\pm$0.1$\%$, showing that the Ramsey contrast decays, and the phase shift is accumulated as a function of $\tau$.

\begin{figure*}
	\begin{center}
		\includegraphics[scale=0.68]{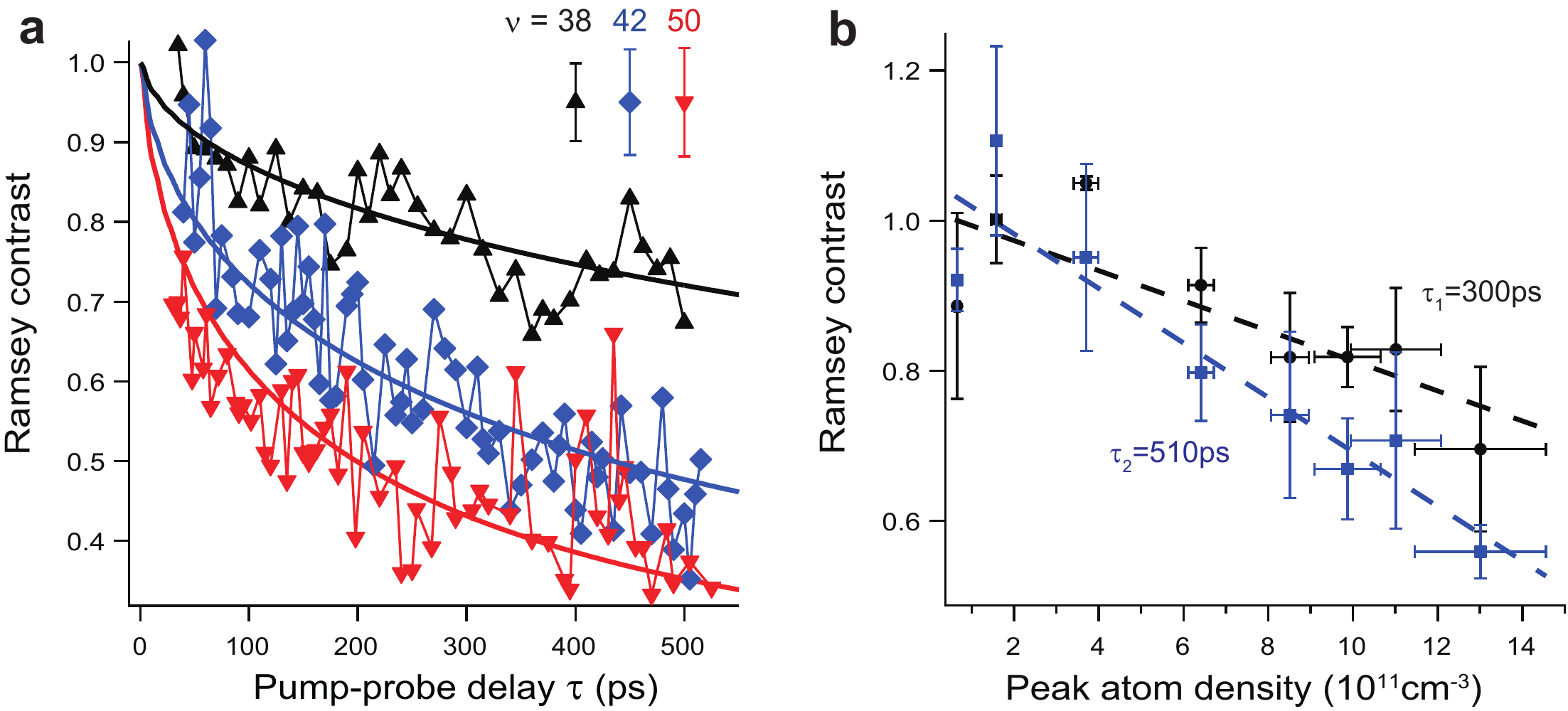}
		\caption{{\bf The principal quantum number and atom density dependences of the Ramsey contrast.}
(\textbf{a}) Measured Ramsey contrasts are plotted as functions of $\tau$ for three different Rydberg levels $\nu=38, 42$ and $50$. The estimated populations $p_\mathrm{e}$ and peak atom densities are $p_\mathrm{e}\sim3.2\,\%$ and $\sim1.2\times10^{12}\,\mathrm{cm}^{-3}$ for $\nu=38$, $p_\mathrm{e}\sim3.3\,\%$ and $\sim1.3\times10^{12}\,\mathrm{cm}^{-3}$ for $\nu=42$, and $p_\mathrm{e}\sim3.1\,\%$ and $\sim1.2\times10^{12}\,\mathrm{cm}^{-3}$ for $\nu=50$, respectively (see Methods sections `Estimation of the atom density' and `Rydberg excitation and detection' for these population and density estimations). The simulations indicated by the black, blue and red solid lines yield adjusting parameters to be $C_6=8, 34$ and $103\,\mathrm{GHz}\,\mu\mathrm{m}^{6}$ for $\nu=38, 42$ and $50$, respectively. The interaction strength in these simulations is limited below $75\,\mathrm{GHz}$, which is the half width half maximum of the pump excitation, and the peak atom density is set to the estimated density for each Rydberg level in these simulations. It should be noted that several Rydberg states are excited for $\nu=50$ (see Supplementary Fig.~4). However, we consider an excitation to a single Rydberg state, to perform the simulations for all of the three Rydberg levels. Each error bar  in the inset represents the average over the error bars (the s.d.) of all data points for each Rydberg level. (\textbf{b}) Ramsey contrast is measured as a function of the peak atom density at two different pump-probe delays $\tau=300$ and $510$\,ps for $\nu=42$ with its population being $\sim3.5\,\%$. The vertical error bars represent the s.d. and the horizontal error bars arise from the density calibration (see Supplementary Note 5 for the estimation of the atom densities plotted in the abscissa).}
		\label{fig04}
	\end{center}
\end{figure*}

It is evident from Fig.~\ref{fig03} that the contrast decay and the phase shift are enhanced when the Rydberg population is increased from $\sim$1.2\,$\%$ to $\sim$3.3\,$\%$.
The strength of the interactions is tuned by varying the Rydberg quantum number $\nu$ and the atom density.
Figure~\ref{fig04}a shows Ramsey contrasts as functions of $\tau$ for three different Rydberg levels $\nu=38$, 42 and 50. 
It is seen from this figure that the contrast decay is accelerated by increasing the principal quantum number $\nu$ of the Rydberg level. The dependence of the Ramsey contrast decay on the atom density is shown in Fig.~\ref{fig04}b, in which the contrast decay is accelerated by increasing the atom density (see Supplementary Note 5 for the estimation of the atom densities plotted in the abscissa of this figure).
From these combined measurements as functions of the Rydberg population, principal quantum number and atom density, we conclude that the observed contrast decay and the phase shift are induced by Rydberg interactions.

The origin of the observed behaviour of the contrast decay and the phase shift is further investigated in Fig.~\ref{fig03}. Here we compare the experimental data with the Ramsey contrast decays and the phase shifts calculated for nearest-neighbour interactions without considering interactions among three or more Rydberg atoms (solid lines). 
The zero-delay offset of the calculated phase shift is arbitrary and adjusted so that the average of the first ten data points is equal to the calculated phase shift averaged over the delay window for those ten data points.
Results for nearest-neighbour interactions are obtained by simple numerical calculations following those performed in previous Ramsey studies \cite{Nipper2012b,Zhou2014}, in which the time-domain Ramsey oscillations modulated by nearest-neighbour interactions are averaged over the distribution of nearest-neighbour distances (see Supplementary Note 6 for details).
Here, the Ramsey oscillation is obtained by solving a Schr\"odinger equation with a Hamiltonian for two interacting atoms.
We assume a van der Waals interaction of the form $U(r)=-C_6/r^6$ with the coefficient $C_6$ being an adjusting parameter.
It is seen in the insets of Fig.~\ref{fig03}a,b that the calculated Ramsey contrast decay becomes slightly faster as the $C_6$ value is increased, but does not grow beyond the decay thresholds 0.988 and 0.967 determined by the Rydberg population $p_\mathrm{e}\sim1.2\,\%$ and $3.3\,\%$, respectively (for details on these thresholds, see Supplementary Note 6).
Similary, the calculated phase shift is almost constant as the $C_6$ value is increased, as shown in Fig.~\ref{fig03}c,d.
These features are not altered by changing the character of the interaction such as van der Waals and dipole-dipole~(see Supplementary Note 6 for details).
The contrast decays and phase shifts observed experimentally are thus clearly larger than expected for nearest-neighbour interactions.
It is therefore concluded that our experimental observation of the Ramsey contrast decay and phase shift on the attosecond timescale demonstrates interactions among more than two Rydberg atoms, and the effect of those interactions on the electron dynamics can be actively controlled by tuning the population and principal quantum number of the Rydberg level, as well as the atom density.

\subsection{Test of a mean-field model}
To model these observations, we first apply a mean-field model (see Supplementary Note 7 for its details).
As in the calculations with nearest-neighbour interactions above, we assume a van der Waals interaction of the form $U(r)=-C_6/r^6$ with the coefficient $C_6$ being the only fitting parameter, which is optimized to reproduce the measured Ramsey contrast in Fig.~\ref{fig03}b by a least-squares fitting.
The outline of the least-squares fitting is given in Supplementary Note 8.
The results of the mean-field simulations are plotted in Fig.~\ref{fig03} (dashed lines).
Figure~\ref{fig03}a,b show that  reasonable agreements can be found for the Ramsey contrast between the measurements and the mean-field simulations with $C_{6} = 1.9\,$GHz$\,\mu$m$^{6}$.
This value for $C_{6}$ yields the phase shifts simulated by the mean-field model in Fig.~\ref{fig03}c,d, where the zero-delay offset is adjusted in the same way as in the calculations with nearest-neighbour interactions above.
Figure~\ref{fig03}c shows that the mean-field simulation yields a phase-shift larger than the measured one by a factor of $\sim$ 4, failing to reproduce our observations.
This discrepancy between measured and simulated phase shifts is not improved by assuming a dipole-dipole interaction,  a hybrid form of a dipole-dipole and a van der Waals interaction or by introducing anisotropic interactions (see Supplementary Figs~2 and 3).
The closer agreement between the measured and simulated phase shifts for $p_\mathrm{e}\sim3.3\,\%$ in Fig.~3d than for $p_\mathrm{e}\sim1.2\,\%$ in Fig.~3c is understood as follows. The Gaussian atom-density distribution of our experimental setup leads to the saturation of the phase shift at longer pump-probe delays. This is because the rapid decrease of the atom density in the Gaussian tails results in the suppression of the contribution to the phase shift from atoms distant from the centre, and therefore the phase shift does not grow afterwards. This saturation is reached both by the measured and simulated phase shifts for $p_\mathrm{e}\sim3.3\,\%$ within our measurement time $500$\,ps (unlike the case with $p_\mathrm{e}\sim1.2\,\%$ and therefore weaker interactions), giving their closer agreement for $p_\mathrm{e}\sim3.3\,\%$ than for $p_\mathrm{e}\sim1.2\,\%$. This saturation effect is explained more quantitatively in Supplementary Note 9.

\subsection{Beyond mean-field analysis}

Next, we apply an exactly solvable theory model~\cite{Feig2013,Hazzard2013,Hazzard2014,Mukherjee2015,Sommer2016} to the observations. Details of this model are presented in ref. 54. 
Briefly, we represent each atom as a two-level system, which is a pseudo-spin system, consisting of a ground state $|\mathrm{g}\rangle$ ($\equiv \left|\downarrow \right\rangle$) and an excited Rydberg state $|\mathrm{e}\rangle$ ($\equiv \left|\uparrow \right\rangle$) with energies $E_\mathrm{g}$ and $E_\mathrm{e}$, respectively.
The experiment consists of four stages: (i) the pump excitation, (ii)  an evolution with an $N$-atom Hamiltonian $H$, (iii) the probe excitation and (iv) the population measurement by field ionization.
The $N$-atom Hamiltonian in stage (ii) includes the atomic energies and the interactions $U(r_{jk})$ between atoms in the Rydberg states as follows:
\begin{eqnarray}
\label{eq:The1}
H = \sum_{j=1}^{N} \hbar \omega \frac{1+\hat{\sigma}^z_j}{2} + \sum_{k=1}^{N-1}\sum_{j>k}^{N} U(r_{jk})\frac{1+\hat{\sigma}^z_{j}}{2}\otimes\frac{1+\hat{\sigma}^z_{k}}{2},
\end{eqnarray}
with $\omega = (E_\mathrm{e} - E_\mathrm{g})/\hbar$ the atomic-resonance frequency and $\hat{\sigma}^z_j$ the Pauli matrix, which is used to represent the internal states of the pseudo-spin at position $j$ (see Methods section `Outline of the exactly solvable model simulation' for more details)~\cite{Feig2013,Hazzard2013,Hazzard2014,Mukherjee2015,Sommer2016}.

\begin{figure}[ht]
	\begin{center}
		\includegraphics[scale=0.7]{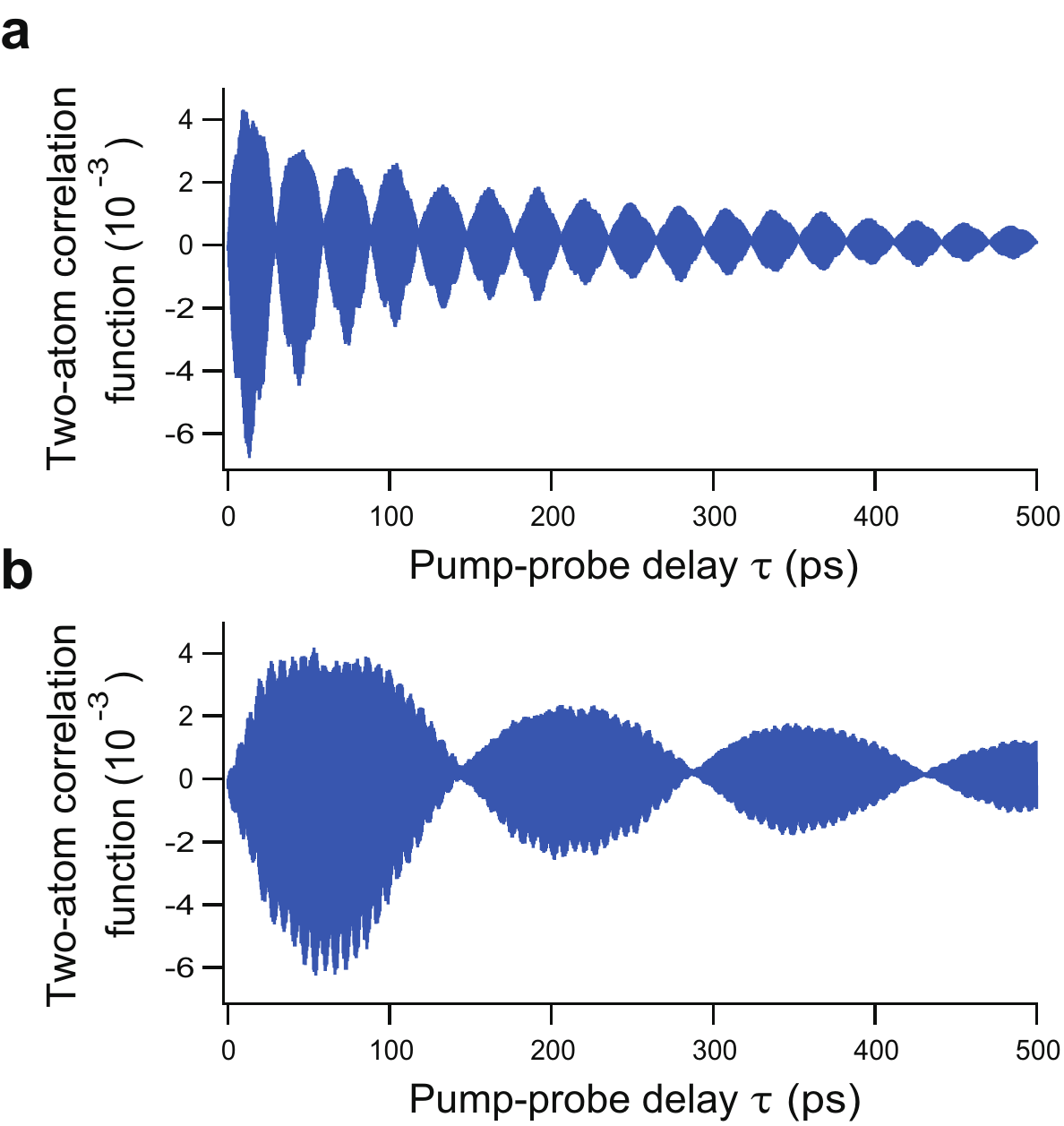}
		\caption{{\bf Two-atom correlation simulated by the exact model.}
The two-particle correlation function is calculated for two atoms in the Gaussian atom density distribution of the present Ramsey measurements for the Rydberg level $\nu=42$. The one atom is located at the centre of the distribution, and the distance to the other atom is set to $1\,\mu$m (\textbf{a}) and $1.3\,\mu$m (\textbf{b}), respectively. The $C_6$ coefficient, the Rydberg population and the peak atom density are set to $34\,\mathrm{GHz}\,\mu\mathrm{m}^{6}$, $\sim3.3$\,\% and $\sim1.3\times10^{12}\,\mathrm{cm}^{-3}$, respectively. The interaction strength is limited below $75\,\mathrm{GHz}$, which is the half width half maximum of the pump excitation.
The coherent superposition of different clusters of interactions leads to dephasing and therefore the global decay.}
		\label{fig05}
	\end{center}
\end{figure}

An exact solution for the time evolution with a Hamiltonian of the form of equation~(\ref{eq:The1}) has recently been presented in refs 50--54 (see also similar numerical approaches in refs 55,56).
This allows for deriving an expression for the exact time evolution of the Ramsey signal $P(\tau)$ for any strength of interactions~\cite{Feig2013,Hazzard2013,Hazzard2014,Mukherjee2015,Sommer2016}. 
For any given atom $j$ interacting with $N-1$ neighbouring Rydberg atoms, one obtaines
\begin{eqnarray}
\label{eq:pop}
P_{j}(\tau)=2p_\mathrm{g}p_\mathrm{e}\Re\Bigg\{ 1 + e^{i(\omega\tau + \phi)}
\prod_{\begin{smallmatrix} k=1 \\ k \neq j\end{smallmatrix}}^{N}\left(p_\mathrm{g} + p_\mathrm{e}e^{i\Delta_{jk}\tau}\right)\Bigg\},
\end{eqnarray}
where $p_\mathrm{g}$ and $p_\mathrm{e}$ are the ground- and Rydberg-state populations, respectively, produced by the initial pump excitation, $\Delta_{jk} = U(r_{jk})/\hbar$ describes a frequency shift induced by the interaction between atoms $j$ and $k$, and $\phi$ is the phase offset arising from the AC-Stark shifts during the picosecond pulse excitations (see Supplementary Note 2).
It is seen from this equation that this model considers different clusters of interactions superposed coherently instead of averaged as performed in our preceding mean-field analysis.
This coherent superposition leads to correlations among different atoms, as exemplified for a two-atom correlation in Fig.~\ref{fig05} (ref. 54).

Further analytical progress is possible by using an approximation, hereafter referred to as a `continuum approximation', in which a continuum function $n({\bf r})$ is considered for the density distribution of Rydberg atoms (see Methods section `Continuum approximation' for details). Briefly, we assume the density distribution to be homogeneous in a small volume around a particular position ${\bf r}$ (ref. 54).
This approximation leads to the following expression  for the Ramsey signal averaged over the whole atomic ensemble
\begin{eqnarray}
\label{eq:The4} P(\tau) = 2p_\mathrm{g}p_\mathrm{e} \left\{ 1 +
|g(\tau)|\cos(\omega \tau+ \alpha(\tau) + \phi)  \right\}.
\end{eqnarray} 
The contrast decay $|g(\tau)|$ and phase shift $\alpha(\tau)$ are then obtained from $g(\tau) \equiv |g(\tau)|e^{i\alpha(\tau)}$,  which, in the case of an isotropic van der Waals interaction and for $\tau > 2\pi/\omega_\mathrm{B}$, is given by
\begin{equation}
\label{eq:The5b} g(\tau) = \frac{2}{\sqrt{\pi}n_\mathrm{p}}\int_{0}^{n_\mathrm{p}} dn \sqrt{\ln\left(\frac{n_\mathrm{p}}{n} \right)}\left[e^{-\sigma_n(1-i)\sqrt{\tau}}\right]e^{\sigma_n\sqrt{\frac{2}{\omega_{\mathrm{B}}\pi}}}.
\end{equation}
Here, $\omega_\mathrm{B} = 2\pi\times 75\,$GHz is the half width half maximum of the pump excitation, $n_\mathrm{p}$ is the peak atom density, the coefficient $2\sqrt{\ln\left(n_\mathrm{p}/n \right)}/(\sqrt{\pi}n_\mathrm{p})$ results from the Gaussian atomic density distribution in the experiment and $\sigma_n =   p_\mathrm{e} n \sqrt{8\pi C_{6}/\hbar}\; (\pi/3)$ is the decay constant~\cite{Sommer2016}.
Equation~\eqref{eq:The5b} predicts that the Ramsey contrast decays approximately as a stretched exponential $e^{-\alpha\sqrt{\tau}}$ with a square-root dependence on the pump-probe delay $\tau$.
This square-root dependence is characteristic of the van der Waals interaction~\cite{Sommer2016}. The decay constant reads $\alpha\approx \sigma_{n_\mathrm{av}}$, where $n_\mathrm{av}= 2^{-3/2}n_\mathrm{p}$ is the average  density for a Gaussian distribution.

\begin{figure*}[ht]
	\begin{center}
		\includegraphics[scale=0.8]{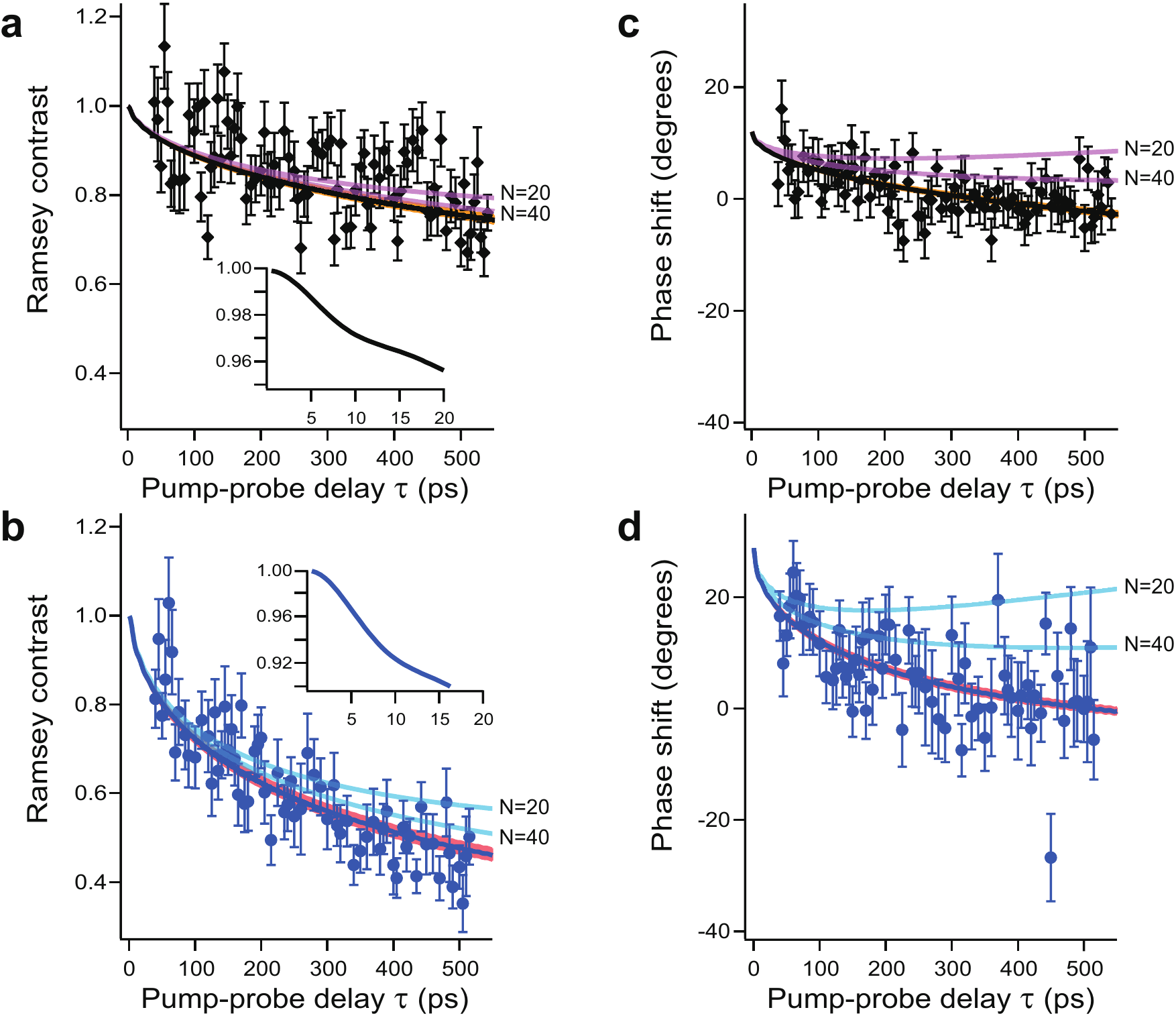}
		\caption{{\bf Measured Ramsey contrasts and phase shifts compared with the numerical ones based on the analytical continuum approximation equation~(\ref{eq:The4}).}
The black diamond-shaped and blue circle data points show the measured Ramsey contrasts (\textbf{a} and \textbf{b}) and the phase shifts (\textbf{c} and \textbf{d}) for the population of the $42$D$_{5/2}$ state being $\sim1.2\,\%$ (\textbf{a} and \textbf{c}) and $\sim3.3\,\%$ (\textbf{b} and \textbf{d}), respectively. In \textbf{a} and \textbf{b}, the measured Ramsey contrasts are compared with the numerical ones based on the analytical continuum approximation for the average number of Rydberg atoms $N=20$ and $N=40$ (purple and blue semi-transparent lines, respectively) as well as for the limit $N\to\infty$ (black and blue solid lines).
The curves for the exact solution agree well with the ones for $N\to\infty$ in the continuum approximation, so that they are indistinguishable on the present scale of the figure. The orange and pink shaded areas correspond to 2 s.d. of the $C_6$ coefficient obtained in the least-squares fitting.
In the insets in \textbf{a} and \textbf{b}, the analytical results in the limit are magnified to show their early quadratic decays at small pump-probe delays originating from the limited bandwidth given by $75\,\mathrm{GHz}$, which is the half width half maximum of the pump excitation. Similarly, the measured and numerical phase-shifts are compared in \textbf{c} and \textbf{d}. The peak atom density is set to $\sim1.3\times10^{12}\,\mathrm{cm}^{-3}$ in these simulations. The error bars represent the s.d.}
		\label{fig06}
	\end{center}
\end{figure*}

\begin{figure}[ht]
	\begin{center}
		\includegraphics[scale=1.0]{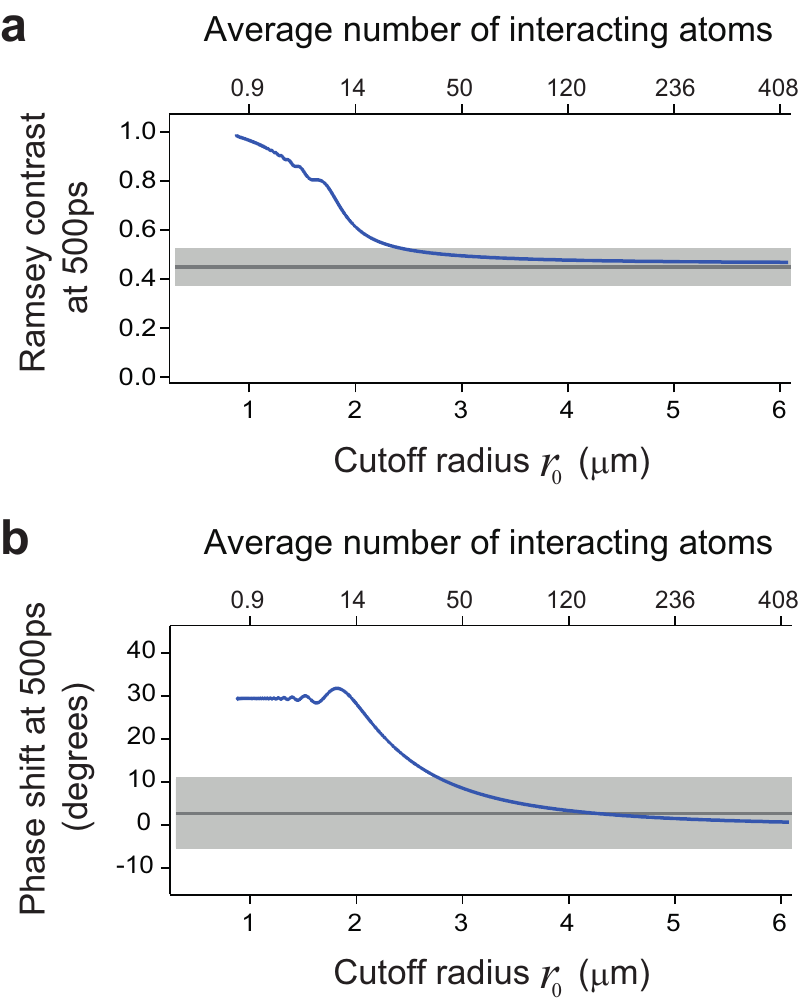}
		\caption{{\bf Convergence of the simulated Ramsey contrast and phase shift as functions of an average number of interacting atoms.} The Ramsey contrast ({\bf a}) and phase shift ({\bf b}) at $\tau=500$\,ps are simulated by the theory model with the continuum approximation and are plotted as functions of the cutoff radius $r_0$ (the lower abscissa) and of an average number of interacting atoms within the volume $V=\frac{4\pi}{3}(r_0^3-r_{\mathrm{B}}^{3})$ (the higher abscissa), where $r_{\mathrm{B}}$ is the blockade radius (see Methods section `Continuum approximation' for more details of $r_0$ and $r_{\mathrm{B}}$).
		The population of the $42$D$_{5/2}$ is set to $\sim3.3\,\%$ in these simulations. The interaction strength is limited below $75\,$GHz, which is the half width half maximum of the pump excitation, and the peak atom density is set to $\sim$1.3$\times10^{12}$\,cm$^{-3}$ in these simulations.
		The results with the van der Waals interaction are displayed by the blue solid lines. The dark-grey solid lines represent the measured Ramsey contrast and the phase shift, each of which is the average over eight points around $\tau=500$\,ps in Fig.~\ref{fig06}b,d. The light-grey shaded area represents 1 s.d. of the average over those eight measured values. Similar results with a dipole-dipole interaction and a hybrid form of a dipole-dipole and a van der Waals interaction are shown in Supplementary Fig.~5.}
		\label{fig07}
	\end{center}
\end{figure}

Figure~\ref{fig06} shows comparisons between (i) the experimental data for the contrast decay and phase shift as functions of $\tau$ for the 42D$_{5/2}$ state and (ii) their numerical results based on the analytical continuum approximation  equation~(\ref{eq:The4}) (solid lines).
The curves for the exact numerical results agree well with the analytical ones for $N \to\infty$, so that they are indistinguishable on the present scale of the figure.
The figure demonstrates that both the exact and analytical results obtained for $C_6=34$\,GHz\,$\mu$m$^6$ agree well with the measured Ramsey contrasts and phase shifts for both of the Rydberg populations $p_\mathrm{e}\sim$1.2\,$\%$ and 3.3\,$\%$ used in the measurement.
Figure \ref{fig06} also shows numerical results with the continuum approximation for the Ramsey contrasts and phase shifts obtained for a finite average number of interacting Rydberg atoms $N=20$ and $40$ (semi-transparent lines; see Methods section `Continuum approximation'). The agreement between the numerical and experimental results improves monotonically with increasing $N$.
Figure~\ref{fig07} shows the Ramsey contrast and phase shift measured (the dark-grey solid lines accompanied by light-grey shaded areas) and simulated with the continuum approximation (the blue solid lines) at the pump-probe delay $\tau\sim500$\,ps.
It is seen from Fig.~\ref{fig07}b that the blue solid line crosses the upper boundary of the light-grey shaded area, which is a confidence interval of the measured value, at an average number of interacting atoms around 40, showing that more than $\sim40$ atoms correlated are necessary to reproduce the measured phase shift.

In the theory model above, the $C_6$ coefficient of the van der Waals interaction serves as the only fitting parameter for the Ramsey contrast in Fig.~\ref{fig06}b. The $C_6$ value thus obtained is used for calculating the contrast decay in Fig.~\ref{fig06}a and the phase shifts in Fig.~\ref{fig06}c,d. Then, the zero-delay offset of the phase shift serves as the only adjustable parameter for the phase shift in each of Fig.~\ref{fig06}c,d.
For distances shorter than the average interatomic separation, the bandwidth of our Rydberg excitation with the picosecond infrared and blue pulses covers multiple adiabatic interaction potentials that can hybridize with the one correlating asymptotically to the $42\mathrm{D}+42\mathrm{D}$ limit (see Supplementary Fig.~1). Although in principle this could affect the contrast decay, the present model based on a single effective potential captures the observed dynamics both for the Ramsey contrast and the phase shift
(see Supplementary Note 10 for the effective treatment of the interaction potentials).

We note that the predicted decay constant for the Ramsey contrast for $p_\mathrm{e}$ $\sim3.3\,\%$ is approximately given by  $\sigma_{n_\mathrm{av}} \sim \sqrt{2\pi}\times1.45\times10^{4}\,(\mathrm{Hz})^{1/2}$. 
The pump-probe delay to reach $e^{-1}(\sim0.37)$ of the initial contrast is thus given by $1/\sigma_{n_\mathrm{av}} ^{2} \sim 760\,$ps, which agrees well with our measurement time of $500\,$ps giving the contrast reduction to $45\,\%$.

We have further employed $g(\tau)$ in equation~(\ref{eq:The5b}) to reproduce the measured Ramsey contrasts shown in Fig.~\ref{fig04}a for three Rydberg levels $\nu=38$, $42$ and $50$, finding good agreement. The phase shifts for these levels have also been measured and analyzed in Supplementary Note 5.
It should be noted that several Rydberg states are excited for $\nu=50$ (see Supplementary Fig.~4). However, the simulation with a single Rydberg state gives good agreement with the experimental results as shown by the solid lines in Fig.~\ref{fig04}a.

We have also performed additional calculations of the measured Ramsey contrasts and phase shifts using dipole-dipole interactions, a hybrid form of a dipole-dipole and a van der Waals interaction, and an anisotropic van der Waals interaction. Although all results are in qualitative agreement with the experimental data, we find that the results for pure van der Waals interactions presented above reproduce the experimental data well with only a single fitting parameter. 
The results of these additional calculations are given in Supplementary Note 10.

\subsection{Origin of the failure of a mean-field model}

The overestimation of the phase shift by the mean-field approximation in Fig.~\ref{fig03}c is intuitively understood as follows. In our actual measurement, the Ramsey contrast of each atom decays due to many different frequencies corresponding to different clusters of interactions superposed coherently, as is expected from equation~(3) of the exact theory model. In the mean-field model, however, the Ramsey oscillation of each atom has its own mean frequency, interacting with `the other atoms' as a whole, so that its contrast does not decay, but the decay arises only from an ensemble average of Ramsey oscillations of many atoms phase shifted from each other. Accordingly, those phase shifts need to be overestimated, as seen in Fig.~\ref{fig03}c, to reproduce the measured contrast decay.

\begin{figure*}[ht]
	\begin{center}
		\includegraphics[scale=0.8]{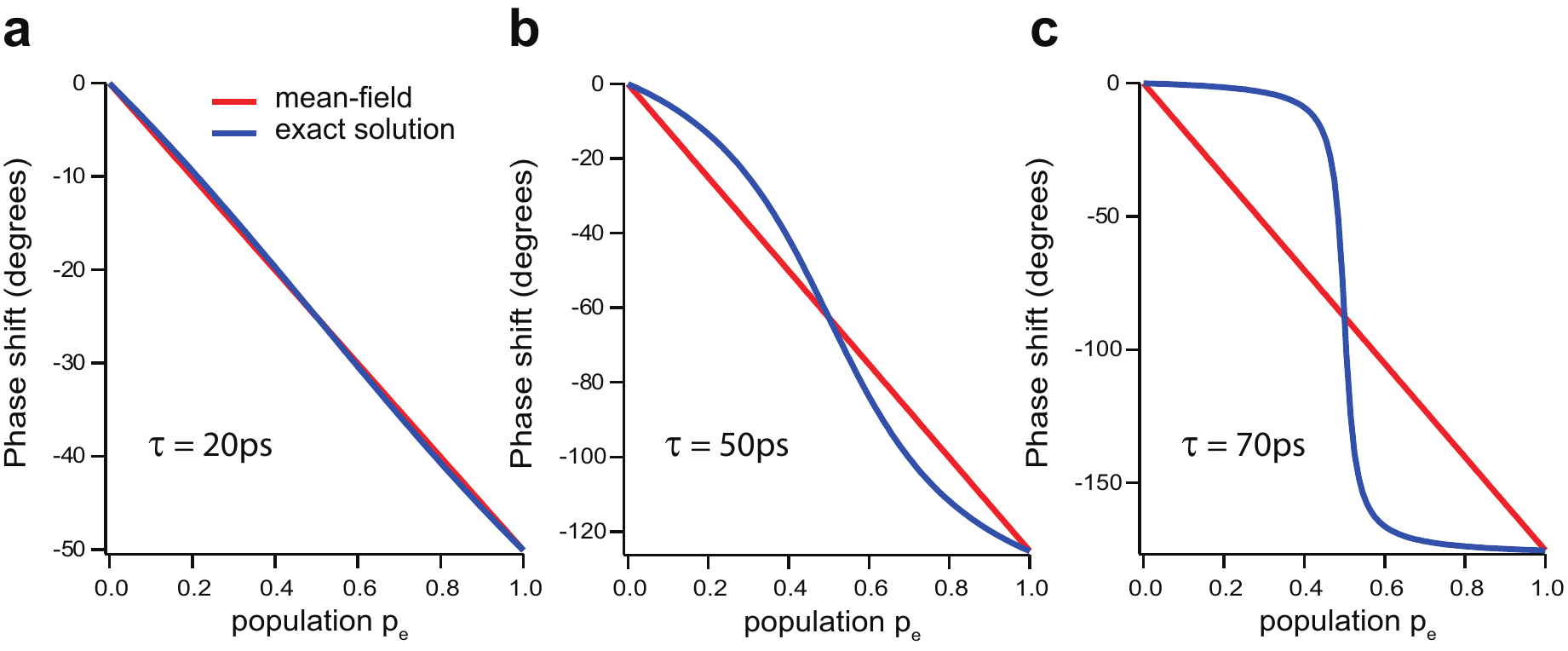}
		\caption{{\bf Comparison between the phase shifts obtained by the mean-field and exact calculations.} In {\bf a} -- {\bf c}, the phase shifts obtained by the mean-field (red solid lines) and exact (blue solid lines) calculations are plotted as functions of the Rydberg population $p_\mathrm{e}$ at the pump-probe delays $\tau=20$, $50$ and $70$\,ps, respectively. The frequency shifts induced by the interactions are set to $U = 6.96$\,GHz, common to all the pairs of atoms. This value corresponds to the energy shift with $C_6 = 34\,\mathrm{GHz}\,\mu\mathrm{m}^6$ at the average internuclear distance in a Gaussian ensemble with its peak density $1.3\times10^{12}$\,cm$^{-3}$.}
		\label{fig08}
	\end{center}
\end{figure*}

Figure~\ref{fig08} shows a comparison between the phase shifts calculated by the mean field and exact models as functions of the Rydberg population $p_\mathrm{e}$ at three different pump-probe delays $\tau = 20$, $50$ and $70$\,ps. It is seen from this figure that the mean-field phase shift is larger than the exact one for $p_\mathrm{e} < 0.5$. It is also seen from this figure that this difference becomes larger as the effective number of interacting atoms becomes larger at longer pump-probe delays. The mean-field model thus overestimates the phase shift for the present $p_\mathrm{e}\sim0.01$ and $0.03$ ($< 0.5$), and $\tau\sim50$ -- $500$\,ps, as is intuitively understood as described in the preceding paragraph.

\section{Discussion}
The experiments presented here bear similarities with recent investigations on coherent spin-exchange dynamics with rotational states of polar molecules trapped in an optical lattice~\cite{Yan2013,Hazzard2014a}. In these experiments, the effect of single-particle decoherence, which proceeds faster than the interaction timescale ($\sim10\,\mathrm{ms}$), has successfully been circumvented by applying spin echo techniques.
In our experiment, on the other hand, the interaction timescale $\sim1\,\mathrm{ns}$ is about three orders of magnitude shorter than the timescale of single-particle decoherence, which is induced mainly by Doppler broadening and is $\sim1\,\mu\mathrm{s}$ at the temperature $\sim$100\,$\mu\mathrm{K}$ estimated for our Rydberg gas (see Methods section `Estimation of the temperature'). The combination of ultrafast and ultracold approaches thus provides an effective pathway for isolating the observation and control of coherent dynamics of a system from its single-particle decoherence processes.

Experiments similar to the ones demonstrated here could in principle be implemented in thermal cells~\cite{Baluktsian2013,Urvoy2015} if one can compensate for the relevant Doppler shift, which causes a phase shift comparable to the interaction-induced phase shift, and can also compensate the effects of atomic motions. 
In the thermal cell at a room temperature, the atoms are not frozen, but move with kinetic energies at that temperature, so  that the motion would lead to additional phase shifts on the order of $2\pi \bar{v} \tau/\lambda \sim 2 \pi \times 0.3$ under the conditions ($\lambda \sim297$\, nm, $\tau \sim500$\,ps, $\bar{v} \sim170$~m/s), where $\lambda$ is the wavelength that corresponds to the energy difference between the ground and Rydberg states.
In our current ultracold measurements, on the other hand, the atoms move only $40$\,pm on average during the same duration, and this distance is shorter than $\lambda$ by four orders of magnitude. This distance is also shorter than the average nearest-neighbour distance in our Rydberg gas by four orders of magnitude (see Methods section `Estimation of the atom density'). Our Rydberg gas is therefore safely regarded as a frozen gas, so that the effects of the atomic motions and collisions on the measurements of coherent dynamics are negligible.

In our present study, a combination of the contrast and phase measurements serves as a useful tool to observe the effects of many-body correlations.
The correlations would be further verified by additional measurements of variance in Ramsey signals, as is performed in refs 8,57, to observe similar many-body correlation effects. Combining a microscope~\cite{Schauss2012,Bloch2015,Browaeys2015} with our experimental setup offers another future possibility to observe many-body correlations more directly in a spatially resolved manner.

We anticipate promising future applications of applying ultrashort coherent laser pulses to ultracold Rydberg gases. Using alternative excitation schemes, one may investigate beyond mean-field effects in Ramsey experiments with more complex Hamiltonians such as Heisenberg-type Hamiltonians of interest for polar molecules~\cite{Yan2013,Hazzard2014a} and atomic clocks~\cite{Martin2013} in optical lattices.
Another application could be the investigation of a scenario in which Rydberg electronic  wave functions are spatially overlapped between neighbouring Rydberg atoms~\cite{Ohmori2014}.
This could lead to new exotic phases in which the Rydberg electrons are shared among many nuclei, and exchange interactions play key roles in their dynamical properties on the ultrafast timescales.
Such a metal-like many-body Rydberg state would naturally lead to Penning ionization quite rapidly. However, Jaksch and colleagues~\cite{Kiffner2015} have theoretically estimated the lifetime of such a metal-like Rydberg state of two $^{85}$Rb atoms ($\nu=50$) to be $\sim100$\,ns. This is longer than our measurement timescale by more than two orders of magnitude, rendering the observation of this state possible using our time-domain approach.

\section{Methods}

\subsection{Atom preparation}
A magneto-optical trap (MOT) of $^{87}$Rb atoms was loaded from background vapour for $1.4\,$s. During the subsequent MOT compression for $30\,$ms, an optical dipole trap was turned on. The dipole trap was composed of a single $1,064\,$nm beam with its power and beam waist being $\sim4\,$W and $\sim30\,\mu$m ($1/e^2\,$radius), respectively. Polarization gradient cooling was performed for $100\,$ms. The trapped atoms were then transferred into the $\mathrm{F}=1$ ground state by switching off the MOT repump laser. After that, we turned off the MOT trapping beams and the magnetic field. While keeping the intensity of the dipole trap laser, plain evaporative cooling was carried out for $50\,$ms.
During the evaporation process, a $76\,\mu$T homogeneous magnetic field was turned on, pointing along the direction of the dipole trap laser. For the next $200\,\mu$s, the atoms were optically pumped to the $|5\mathrm{S}_{1/2}, \mathrm{F}=2, \mathrm{m}_{\mathrm{F}}=+2\rangle$ state by using the MOT repump beam and a $\sigma^+$ beam, which is resonant to the transition from $|5\mathrm{S}_{1/2}, \mathrm{F}=2\rangle$ to $|5\mathrm{P}_{3/2}, \mathrm{F}'=2\rangle$ and counterpropagates with the dipole trap laser. The dipole-trap laser was turned off $2\,\mathrm{\mu s}$ before the irradiation of the picosecond infrared and blue pulses to avoid $2$+$1$ multiphoton ionization induced by a combination of the picosecond pulses and the trapping laser beam, whereas the homogeneous magnetic field remained on.
The picosecond infrared and blue pulses at $\sim779$ and $\sim481\,$nm, which propagated collinearly with the dipole trap beam, had cross-sections with FWHMs of $\sim130\,(100)\,\mu$m and $30\,(30)\,\mu$m along the $x$ ($y$) direction (Fig.~1a).

\subsection{Estimation of the atom density}
At first, the atom density in the Ramsey measurements was estimated solely from the total number of atoms and the size of the atomic ensemble obtained by {\it in-situ} absorption imaging with a CCD (charge-coupled device) camera without expanding the atomic ensemble. However, in contrast to the axial size of the atomic ensemble ($\sim2\,$mm FWHM), the spatial resolution of the {\it in-situ} absorption imaging with the CCD camera was not high enough mostly because of the aberrations of the imaging lens to resolve the radial size, resulting in an underestimation of the atom density.

In a later independent experiment, hereafter referred to as a `reference experiment', we estimated the radial size from the temperature of the atomic ensemble and the trap frequency of the radial direction, setting the trapping conditions almost the same as those employed in the Ramsey measurements.
Those trapping conditions set almost the same were (1) the same loading sequence as described in the preceding subsection; (2) the dipole-trap laser power (with an error described in the next sentence); (3) the dipole-trap laser focusing; and (4) the total number of atoms (with an error described in the next sentence).
We performed two reference experiments under two different trapping conditions corresponding to the higher and lower densities in the Ramsey measurements, setting the dipole-trap laser power (the trapping condition (2)) with a difference of $\sim2\,\%$ and $\sim5\,\%$ from the power averaged over the Ramsey measurements for the higher and lower densities, respectively, and setting the total number of atoms (the trapping condition (4)) within 1 s.d. $\sim14\,\%$ and $\sim33\,\%$ of the eight and seven corresponding values measured in the Ramsey measurements for the higher and lower densities, respectively.
In these reference measurements, the temperature was measured by an expansion of the atomic ensemble and parametric heating was employed to infer the radial trap frequency.
The temperatures and the trap frequencies were thus obtained to be $\sim67\,\mu$K and $2.2$\,kHz and $\sim39\,\mu$K and $1.1$\,kHz under those two trapping conditions, respectively.
These temperatures and trap frequencies gave typical radial sizes (FWHM) of the atomic ensembles in the reference experiments to be $\sim14$ and $20\,\mu$m, which we regarded to be the radial sizes of the atomic ensembles in the Ramsey measurements for the higher and lower densities, respectively.

The total number of atoms was also slightly underestimated by the {\it in-situ} absorption imaging in the Ramsey measurements because of the spatial resolution, so that it was calibrated in a later independent experiment in which the loading sequence and the trap laser focusing (the trapping conditions (1) and (3)) were almost the same as those employed in the Ramsey measurements.
In this calibration experiment, we measured the total number of atoms by the absorption imaging with the expansion of the atomic ensemble as a function of the one measured without the expansion to obtain a linear calibration curve with its slope being $1.04\pm0.03$. This linear calibration curve gave the total numbers of atoms to be $\sim6\times 10^5$ and $\sim4\times 10^4$ for the higher and lower densities in the Ramsey measurements, respectively, for $\nu=42$.

These radial sizes and the total numbers of atoms were combined with the axial sizes measured {\it in situ} in the Ramsey measurements, to give the peak atom densities of $\sim1.3\times10^{12}$ and $\sim4\times10^{10}\,\mathrm{cm}^{-3}$ for the higher and lower densities, respectively, for $\nu=42$.
Similarly, the peak atom density and the total number of atoms for $\nu=38$ were estimated to be $\sim1.2\times10^{12}\,\mathrm{cm}^{-3}$ and $\sim5\times 10^5$ for the higher density and  $\sim4\times10^{10}\,\mathrm{cm}^{-3}$ and $\sim4\times 10^4$ for the lower density, respectively, and for $\nu=50$ they were estimated to be $\sim1.2\times10^{12}\,\mathrm{cm}^{-3}$ and $\sim4\times 10^5$ for the higher density and  $\sim3\times10^{10}\,\mathrm{cm}^{-3}$ and $\sim3\times 10^4$ for the lower density, respectively.

\subsection{Estimation of the temperature}
The temperature of the atomic ensemble was not measured {\it in situ} in a series of the Ramsey measurements; however, it was measured in a later independent experiment, hereafter referred to as a `temperature experiment', in which we set the trapping conditions (1)—(4) in the same way as described in the preceding subsection. In this temperature experiment, the temperature was measured by an expansion of the atomic ensemble. Two temperature experiments were performed independently under the trapping conditions corresponding to the higher density in the Ramsey measurements, giving $\sim67$ and $\sim72\,\mu$K, respectively, so that we estimated the temperature of the higher density ensemble to be $\sim70\,\mu$K. Similarly, a temperature experiment was performed independently under the trapping conditions corresponding to the lower density in the Ramsey measurements, giving $\sim 39\,\mu$K, so that we estimated the temperature of the lower density ensemble to be $\sim39\,\mu$K.

\subsection{Rydberg excitation and detection}

The output of a Ti:sapphire laser system (Spectra Physics; Spitfire Ace, wavelength $\sim779\,$nm, pulse width $\sim1\,$ps, repetition rate $1\,$kHz) was used as the infrared pulse and was also used to pump an optical parametric amplifier (Spectra Physics; TOPAS) to generate the blue pulse tuned to $\sim 481\,$nm. The repetition rate was reduced using pulse pickers to synchronize it with the atom preparation sequence whose duration was $1.6$\,s. To reduce the number of Rydberg states to be excited, the spectra of those infrared and blue pulses were cut using homemade pulse shapers in a 4$f$-setup ($f=500\,$mm), respectively. The infrared and blue pulses were combined collinearly with a dichroic mirror and their relative timing was coarsely adjusted to be zero by cross-correlation measurements based on sum frequency generation and was further optimized to maximize the Rydberg ion signals. They were introduced into a Michelson-type interferometer to produce a pair of identical double pulses, each of which was composed of the infrared and blue pulses. Those two double pulses induced the pump and probe excitations, respectively (see Fig.~\ref{fig01}a). The relative phase between these two double pulses were tuned with attosecond precision~\cite{Katsuki2013}. Those two double pulses were combined collinearly with the dipole-trap laser beam with another dichroic mirror. Those two double pulses and the trapping laser beam were focused with a plano-convex lens ($f=250\,$mm) to the atomic ensemble.

The infrared and blue pulses, and the optical pumping beam were circularly polarized in the same direction with respect to the magnetic field, so that the state $\nu \mathrm{D}_{5/2}, \mathrm{m}_{\mathrm{J}}=+5/2$ was mostly populated, and excitations to S Rydberg states were suppressed due to transition selection rules even though the effective two-photon excitation spectrum covered the S states. This excitation scheme also suppressed the Raman transition between the F=2 and F=1 hyperfine states in the ground state that could happen within the single infrared pulse to induce undesirable beating in the Ramsey contrast as a function of $\tau$, of which an example is shown in Fig.~\ref{fig02}d, with a period of $146\,$ps, which is the reciprocal of the hyperfine splitting $6.83\,$GHz.

The maximum population of the $\nu$D Rydberg states was not $>5\,\%$, to suppress photo-ionization. The typical pulse energies of the infrared and blue pulses were $\sim10$ and $400$\,nJ, respectively, for the $1.2\,\%$ population and were $\sim30$ and $600$\,nJ, respectively, for the $3.3\,\%$ population. The population was estimated from the loss of the number of the ground-state atoms in the higher-density ensemble induced by the irradiation of the picosecond infrared and blue pulses. We measured the number of the atoms by absorption imaging with and without the picosecond pulses and compared those numbers to evaluate the loss.
See the previous subsection `Estimation of the atom density' for the details of the evaluation of the atom number by absorption imaging.
The loss induced by a single pair of the infrared and blue pulses was too small (a few percent) to be evaluated securely, so that we shined $30$ or $50$ pairs of the infrared and blue pulses, each of which was accompanied by field ionization~\cite{Gallagher1994}, at a $1\,$kHz repetition rate to induce the loss that was clearly visible and was used to infer the loss induced by the single pair of the infrared and blue pulses. The loss for different amounts of pulse pairs followed the relation $(N_\mathrm{r}/N_\mathrm{t})=(1-p_\mathrm{e})^q$, where $N_\mathrm{r}$ is the number of atoms remaining after $q$ pulses and $N_\mathrm{t}$ is the total atom number. $p_\mathrm{e}$ is the Rydberg state population for the atoms in the ensemble.

After the probe excitation, the populated Rydberg states were ionized by means of field ionization. The Rb$^+$ ions thus produced were detected with a micro channel plate placed $5.5\,$cm away from the atomic ensemble. The electric field for the ionization was triggered $50\,$ns after the probe excitation, reaching the ionization threshold within the next $100\,$ns.
The output of the micro channel plate was amplified with a preamplifier and sent to a gated integrator, whose output was fed into a computer. It is noteworthy that the field-ionization spectrum shown in Fig.~\ref{fig01}c was measured using an oscilloscope with the electric field ramped up slowly on the microsecond timescale to check how many Rydberg states were populated by our broadband excitation with the picosecond pulses.

\subsection{Time-domain Ramsey interferometry}
The interaction among Rydberg atoms was observed by time-domain Ramsey interferometry with a pair of two-photon excitations: pump and probe. Their delay $\tau$ was coarsely tuned on the picosecond timescale with a motorized mechanical stage placed in one arm of the interferometer mentioned above and was scanned finely on the attosecond timescale with a piezoelectric transducer to measure Ramsey interferograms. A He-Ne laser beam was introduced to the interferometer to check the linearity of the scan by monitoring its optical interference. The period of this optical interference was also used for the calibration of the pump-probe delay $\tau$. We measured the field-ionization signals of the two atomic ensembles with different densities alternately to suppress systematic uncertainties, scanning $\tau$ in steps of $\sim30\,$as over a range of $\sim3\,$fs at each coarse delay tuned by the mechanical stage on the picosecond timescale.
The obtained Ramsey interferogram was fitted with a sinusoidal function. As the expected energy shift induced by the interaction was at most on the order of $10\,$GHz, much smaller than the eigenfrequency of the Rydberg state itself ($\sim1\times10^{15}\,$Hz), we fitted the interferograms with the same eigenfrequencies for the higher and lower densities to evaluate the phase shift of the higher-density ensemble from the lower-density one. We have defined the contrast of the interferogram to be the ratio of the amplitude of the fitted sinusoidal function to its mean value. In Fig.~\ref{fig02}a--c, the signal intensities are normalized by the mean value of the sinusoidal function fitted to each interferogram. We have defined Ramsey contrast to be the ratio of the contrast of the higher-density ensemble to that of the lower-density one, as shown in Figs~\ref{fig03},~\ref{fig04} and~\ref{fig06}. We assume that the interactions are negligible in the lower-density ensemble, which is thus taken to be a reference for measuring the contrast decay and the phase shift that may be induced by the interactions in the higher-density ensemble.

\subsection{Outline of the exactly solvable model simulation}

We  represent each atom as a two-level system, which is a pseudo-spin system, consisting of a ground state $|\mathrm{g}\rangle$ ($\equiv \left|\downarrow \right\rangle$) and an excited Rydberg state $|\mathrm{e}\rangle$ ($\equiv \left|\uparrow \right\rangle$) with energies $E_\mathrm{g}$ and $E_\mathrm{e}$, respectively.
This assumption should be reasonable, as the contribution of the neighbouring Rydberg levels is small for the $42$D$_{5/2}$ state, as seen from the field-ionization spectrum shown in Fig.~\ref{fig01}c. The $N$-atom wave function $|\Psi(\tau)\rangle_{N}$ is initially assumed to be a product of independent single-atom wave functions $|\Psi(0)\rangle_{N}=|\mathrm{g}\rangle^{\otimes N}$. The experiment consists of four stages: (i) the pump excitation, (ii)  an evolution with an $N$-atom Hamiltonian $H$, (iii) the probe excitation and (iv) the population measurement by field ionization. The experimental observable is the number of Rydberg excited atoms detected as a time-dependent signal $P(\tau) =\langle \Psi(\tau)|_N \hat P |\Psi(\tau)\rangle_{N}$, which is the expectation value of the sum of projection operators  $\hat P =\sum_{j=1}^{N}\hat P_j$, where $\hat P_j \equiv \left| \uparrow \right\rangle_{j}\left\langle \uparrow\right|_{j}/N$ measures the population in the Rydberg state for atom $j$ normalized to the total atom number $N$.

The $N$-atom Hamiltonian in stage (ii) includes the atomic energies and the interactions $U(r_{jk})$ between atoms in the Rydberg states. This is justified on the picosecond timescale of the current measurement, as Rydberg interactions provide coupling strengths on the order of GHz for micrometre separations, much larger than those between Rydberg- and ground-state atoms and between two ground-state atoms on the order of kilohertz and less than kilohertz, respectively, for micrometre separations. This leads to a diagonal Hamiltonian of the form in equation~(\ref{eq:The1}). 

\subsection{Continuum approximation}

In this approximation we assume the density distribution to be homogeneous in a small volume around a particular position ${\bf r}$ (ref. 54).
The Ramsey signal $P({\bf r},\tau)$ is calculated for such a homogeneous region
as
\begin{eqnarray}
\label{eq:method1}
P({\bf r},\tau) &\approx& 2p_\mathrm{g}p_\mathrm{e}\Re\left\{ 1+ e^{i(\omega\tau + \phi)}\left( p_\mathrm{g} + p_\mathrm{e}\gamma(\tau)\right)^{N_{0}({\bf r})-1} \right\},\nonumber \\
\end{eqnarray}
where $\gamma(\tau)=\frac{3}{r_{0}^{3}-r_{\mathrm{B}}^{3}} \int_{r_{\mathrm{B}}}^{r_{0}}dr r^{2} e^{i\frac{U(r)}{\hbar}\tau}$ and $r_0$ is referred to as a cutoff radius, so that we consider interactions only among the atoms within a sphere whose radius is $r_0$. A blockade radius $r_{\mathrm{B}}$ is determined by the bandwidth of the pump excitation and is $\sim0.88\,\mu\mathrm{m}$ in the present experiment. $N_{0}({\bf r})$ is the number of atoms in the local volume $\frac{4\pi}{3} (r_{0}^{3}-r_{\mathrm{B}}^{3})$ and depends on the local density $n({\bf r})$ due to $N_{0}({\bf r}) =  \frac{4\pi}{3} (r_{0}^{3}-r_{\mathrm{B}}^{3})n({\bf r})$.
This signal is averaged over the whole ensemble whose atom-density distribution is assumed to be Gaussian with the peak atom-density $n_\mathrm{p}$ to obtain equation~\eqref{eq:The4}, in which the Ramsey contrast and phase shift can be derived from 
\begin{eqnarray}
\label{eq:method3}
g(\tau) &=& \frac{2}{\sqrt{\pi}n_\mathrm{p}}\int_{0}^{n_\mathrm{p}} dn \sqrt{\ln\left(\frac{n_\mathrm{p}}{n} \right)}\left( p_\mathrm{g} + p_\mathrm{e}\gamma(\tau)\right)^{N_{0}(n)-1}\nonumber \\
\end{eqnarray}
to be the absolute value and phase of the function $g(\tau)$, respectively. In the case that the potential is given by a van der Waals interaction and in the limit $N_{0}(n)\to \infty$, the continuum approximation leads to the expression in equation~(\ref{eq:The5b}) for the function $g(\tau)$.
We expect this approximation to describe the system well for a sufficiently low population of Rydberg excited atoms $p_\mathrm{e} \ll p_\mathrm{g}$~(ref. 54), which is safely satisfied in our experiments.


\vspace{\baselineskip}

\section{Acknowledgements}
We thank T. Pfau, R. L\"ow, D. Jaksch, M. Kiffner, A. Browaeys, T. Lahaye, V. Vuleti\'c, I. Lesanovsky, R.J. Levis, R.J.D. Miller, I. Ohmine, S. Saito, T. Yanai, S. Takeda, A. Tanaka and D. Vodola for fruitful discussions.
We also acknowldge K.R.A. Hazzard and R. Mukherjee for showing us their unpublished numerical simulations of the contrast decays and phase-shifts based on the exact expression equation~(3).
This work was partly supported by CREST-JST, JSPS Grant-in-Aid for Specially Promoted Research Grant Number 16H06289 and Photon Frontier Network Program by MEXT.
C.G. was supported from the Austrian Science Fund (FWF) via project P24968-N27.
G.P. was supported by ERC-St Grant ColdSIM (Number 307688), EOARD, RySQ, UdS via IdEX and ANR via BLUESHIELD.
M.W. was supported by the EU FET Network `Rydberg Quantum Simulators RySQ' and Heidelberg Center for Quantum Dynamics.
K.O. thanks Alexander von Humboldt foundation, University of Heidelberg and University of Strasbourg for supporting this international collaboration.

\clearpage
\onecolumngrid

\renewcommand{\figurename}{{\bf Supplementary Fig.}}
\setcounter{figure}{0}
\setcounter{equation}{0}
\setcounter{section}{0}

\section*{{\Large Supplementary Materials}}

\noindent
{\bf {\Large Supplementary Figures}}\\

\begin{figure}[ht]
	\begin{center}
		\includegraphics[scale=0.7]{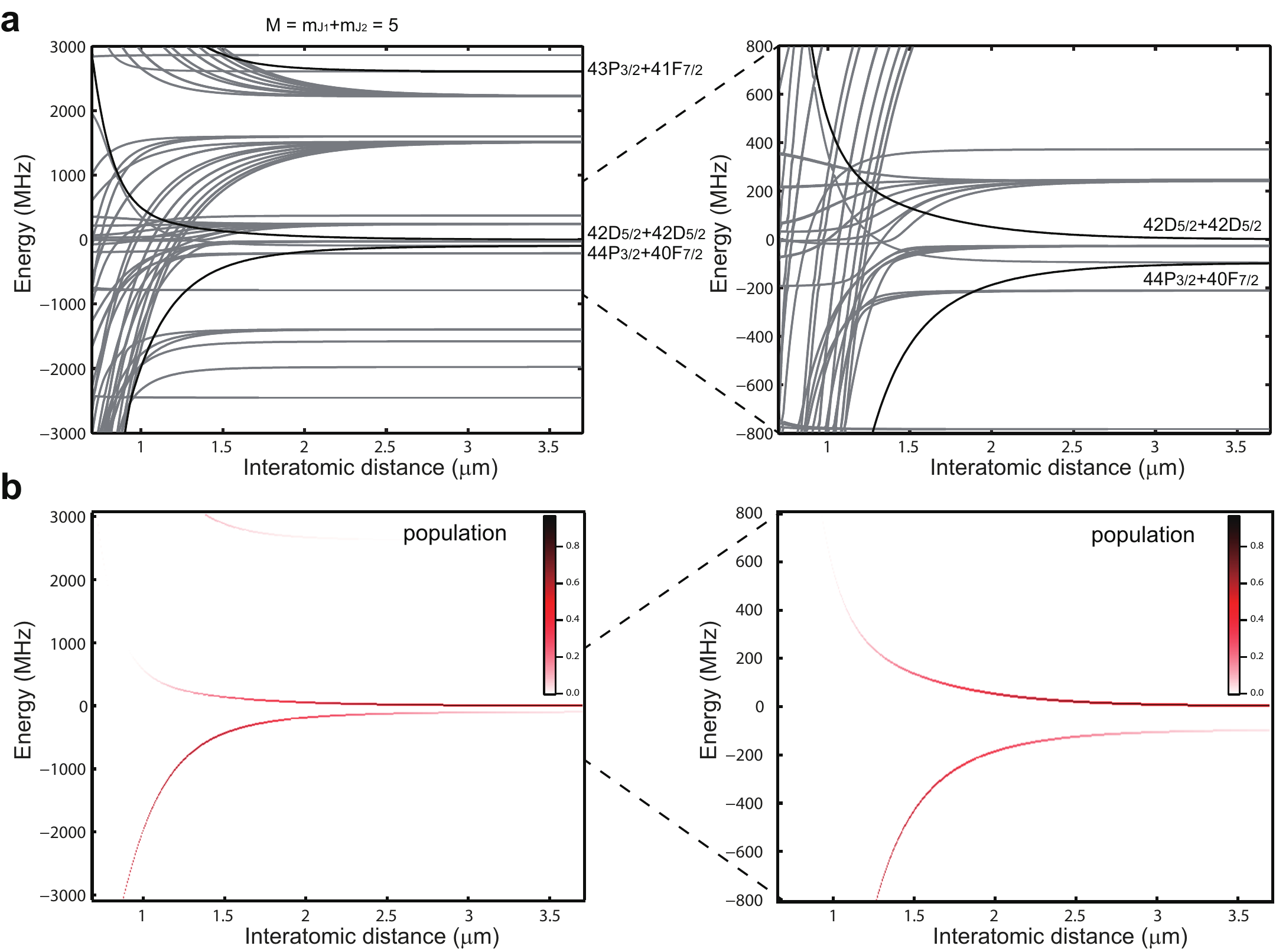}
		\caption{{\bf Two-atom dipolar potentials around the 42D$_{5/2}$ + 42D$_{5/2}$ asymptote.} $\theta = 0$ and $M = m_{\mathrm{J}_{1}} +  m_{\mathrm{J}_{2}} = 5$, where $\theta$ is the same angle as has been introduced in Supplementary Eq.~(\ref{eq:Aniso1}). Only the asymptotic states with electronic angular-momenta below $l = 5$, principal quantum-numbers ranging from 38 to 46, and within $\pm$30\,GHz from the 42D$_{5/2}$ + 42D$_{5/2}$ asymptote are considered in the diagonalization of the potential-energy matrix. ({\bf a}) The potentials that dominate the interaction between two $|$42D$_{5/2}$, m$_{\mathrm{J}}=5/2\rangle$ atoms are indicated by black solid lines. ({\bf b}) The color code indicates the fraction of the population of the asymptotic $|$42D$_{5/2}$, m$_{\mathrm{J}}=5/2\rangle \otimes |$42D$_{5/2}$, m$_{\mathrm{J}}=5/2\rangle$ state that is contained in an interaction-induced mixed state as a function of the interatomic distance. Only the contribution of the $|$42D$_{5/2}$, m$_{\mathrm{J}}=5/2\rangle \otimes |$42D$_{5/2}$, m$_{\mathrm{J}}=5/2\rangle$ asymptotic state needs to be considered since it is the only state in the displayed energy range that can be addressed by the pulse excitation due to selection rules.}
		\label{Suppfig08}
	\end{center}
\end{figure}


\begin{figure}[t]
	\begin{center}
		\includegraphics[scale=0.8]{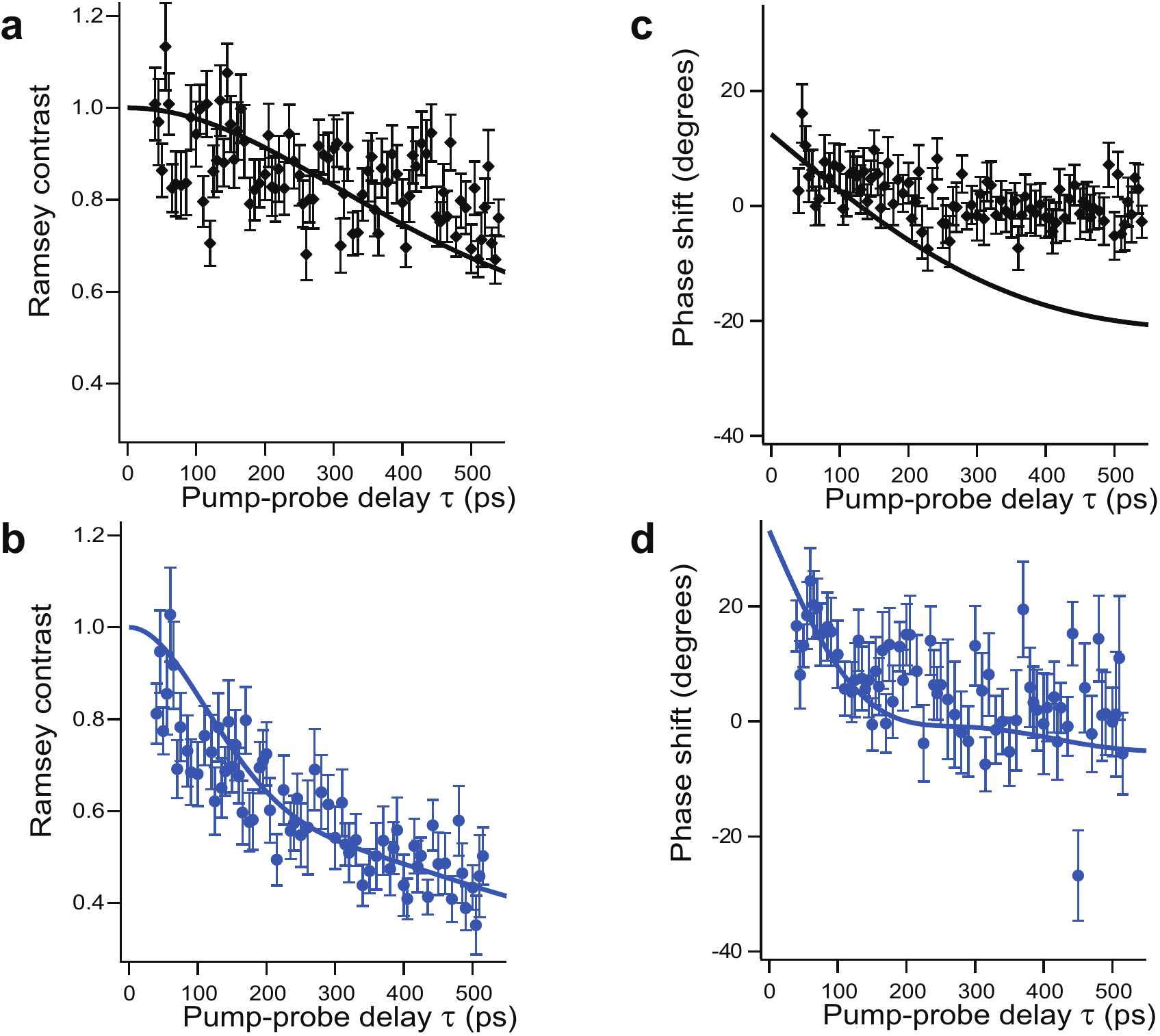}
		\caption{{\bf Mean-field  analysis of the Ramsey contrast and phase-shift with an anisotropic van der Waals potential.} The black-diamond-shaped and blue-circle data-points show the Ramsey contrasts (\textbf{a} and {\bf b}) and the phase-shifts (\textbf{c} and {\bf d}) measured with the population of the $42$D$_{5/2}$ state being $\sim$1.2$\%$ (\textbf{a} and {\bf c}) and $\sim$3.3$\%$ (\textbf{b} and {\bf d}), respectively. In \textbf{a} and \textbf{b}, the Ramsey contrasts simulated by the mean-field model with the anisotropic potential given by Supplementary Eq.~(\ref{eq:Aniso1}) (black and blue solid lines) are compared with the measured ones. Similarly the measured and simulated phase-shifts are compared in \textbf{c} and \textbf{d}. The interaction strength is limited below $75\,$GHz, which is the bandwidth (half width half maximum) of the pump excitation, and the peak atom density is set to $\sim1.3\times10^{12}$\,cm$^{-3}$ in these simulations. The coefficient $C_{6} = 3.4\,$GHz\,$\mu$m$^{6}$ has been used in these simulations. The error bars represent the standard deviation.}
		\label{Suppfig04}
	\end{center}
\end{figure}


\begin{figure}[H]
	\begin{center}
		\includegraphics[scale=0.8]{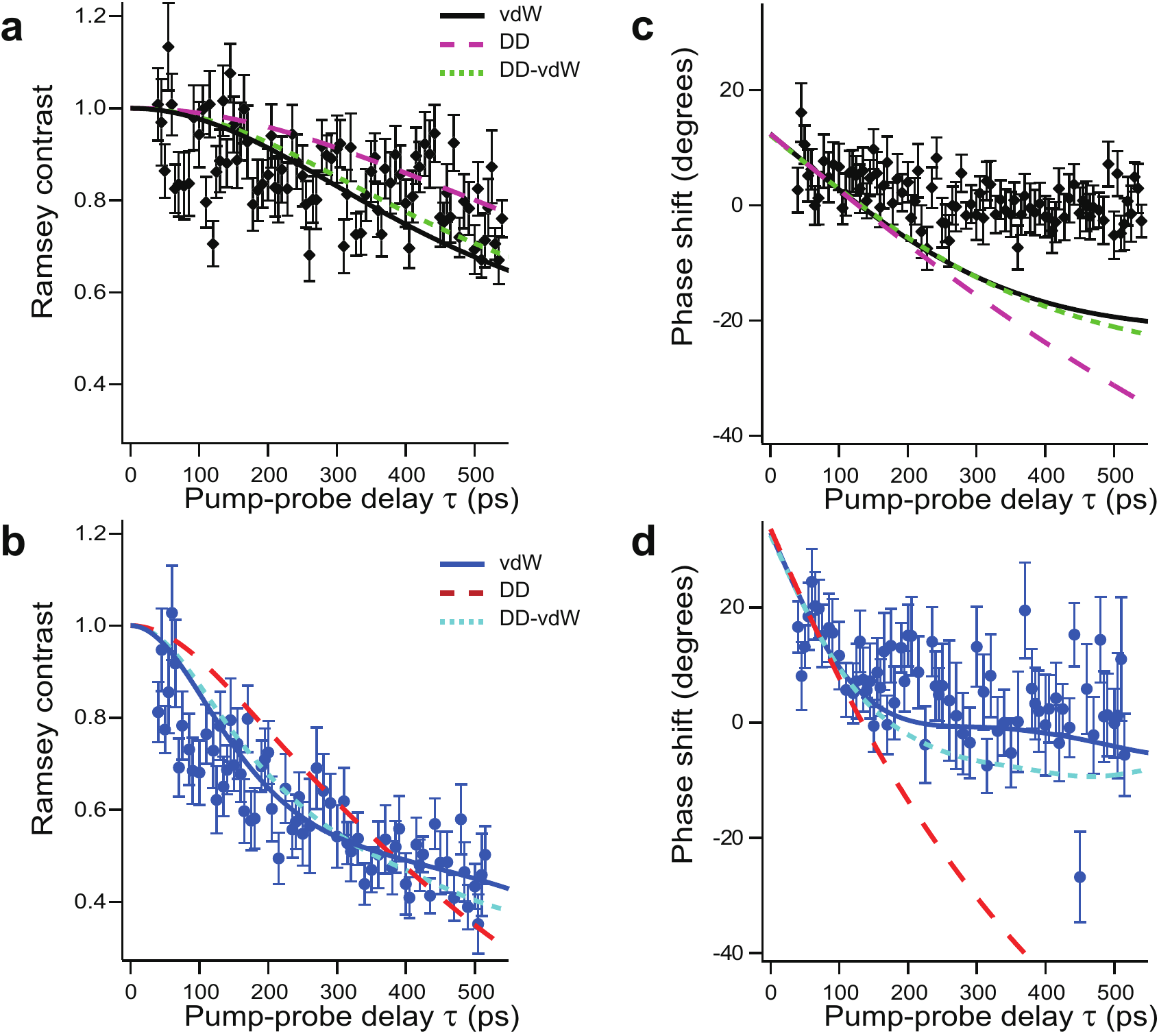}
		\caption{{\bf The mean-field analysis of the Ramsey contrast and phase-shift with the dipole-dipole (DD) interaction and the hybrid form of a dipole-dipole and a van der Waals (DD-vdW) interaction without anisotropies.} The black-diamond-shaped and blue-circle data-points show the Ramsey contrasts (\textbf{a} and {\bf b}) and the phase-shifts (\textbf{c} and {\bf d}) measured with the population of the $42$D$_{5/2}$ state being $\sim$1.2$\%$ (\textbf{a} and {\bf c}) and $\sim$3.3$\%$ (\textbf{b} and {\bf d}), respectively. In \textbf{a} and \textbf{b}, the Ramsey contrasts are simulated by the mean-field model with the DD (dahed line) and DD-vdW (dotted line) interactions without anisotropies and are compared with the measured ones. Similarly the measured and simulated phase-shifts are compared in \textbf{c} and \textbf{d}. The mean-field simulations with the pure van der Waals (vdW) interaction (solid line), which have been shown in Fig.~3 in the main text, are presented again to be compared with the DD and DD-vdW result. The interaction strength is limited below $75\,$GHz, which is the bandwidth (half width half maximum) of the pump excitation, and the peak atom density is set to $\sim1.3\times10^{12}$\,cm$^{-3}$ in these simulations. The coefficient $C_{3} = 1\,$GHz\,$\mu$m$^{3}$ has been used for the DD interaction, and a combination of $C_{3} = 3.4\,$GHz\,$\mu$m$^{3}$ and $r_\mathrm{c} = 0.81\,\mu$m have been used for the DD-vdW interaction~[see Supplementary Eq.~\eqref{eq:Eff1}]. The error bars represent the standard deviation.}
		\label{Suppfig06}
	\end{center}
\end{figure}


\begin{figure}[ht]
	\begin{center}
		\includegraphics[scale=1.0]{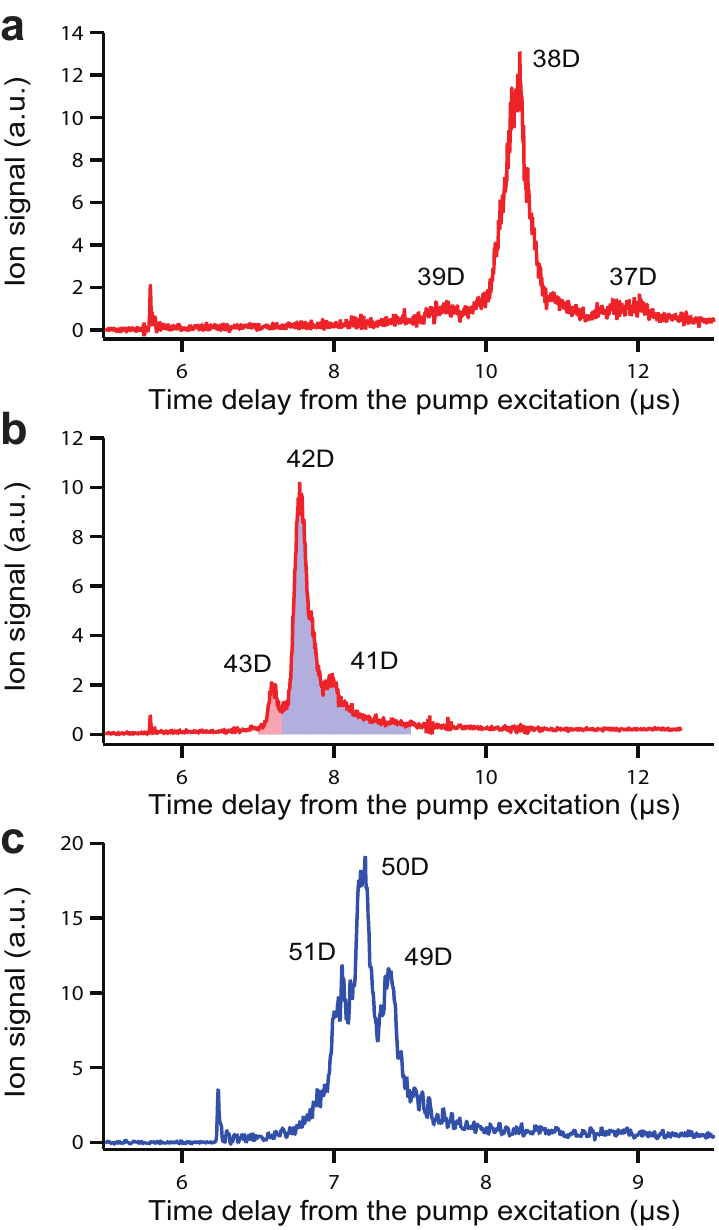}
		\caption{{\bf State-resolved field-ionization spectra.} (\textbf{a}) Ion signal measured as a function of time-delay from the ps pulsed-laser excitations to Rydberg states $\nu=$ $38$. The electric field was ramped at the same speed as the one for Fig.~1c in the main text. The atom density and Rydberg population were  $\sim 4\times 10^{10}\,$cm$^{-3}$ and $3.2\pm0.1$\,\%, respectively. (\textbf{b}) The ion signal for $\nu=42$ with the atom density and Rydberg population being $\sim4\times 10^{10}\,$cm$^{-3}$ and $1.2\pm0.1$\,\%, respectively. This is the same signal as shown in Fig.~1c. The red-shaded region indicates the integration range used for the estimation of the relative population in the 43D state, whereas the blue-shaded area shows the integration range for the 42D and 41D states. (\textbf{c}) The ion signal for $\nu=50$ with the atom density and Rydberg population being $\sim3\times 10^{10}\,$cm$^{-3}$ and $3.1\pm0.2$\,\%, respectively. The ramp-up  speed of the electric field was slower than that for $\nu=42$ and $38$ by a factor of 2/3. The bandwidths of the excitations in these measurements in {\bf a} -- {\bf c} are the same as the one for the pump excitation in the Ramsey measurements.} 
		\label{Suppfig10}
	\end{center}
\end{figure}


\begin{figure}[ht]
	\begin{center}
		\includegraphics[scale=1]{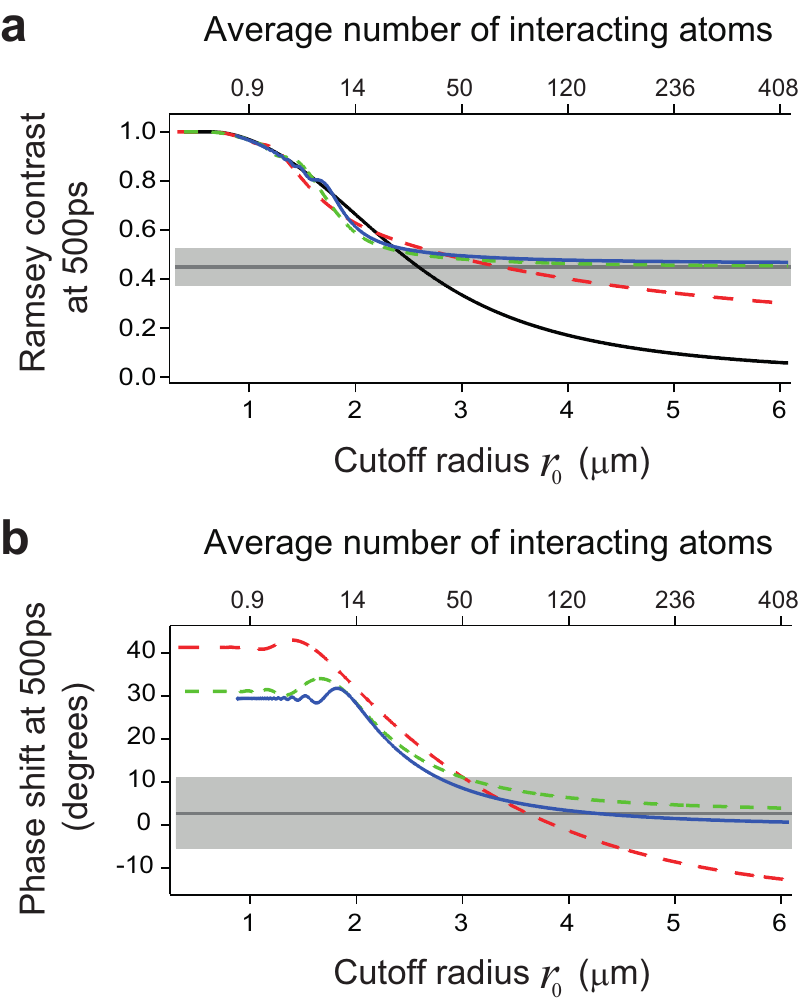}
		\caption{{\bf Convergence of the simulated Ramsey contrast and phase-shift as functions of an average number of interacting atoms.} The Ramsey contrast ({\bf a}) and phase-shift ({\bf b}) at $\tau=$ 500\,ps are simulated by the theory model with the continuum approximation and are plotted as functions of the cutoff radius $r_0$ (the lower abscissa) and of an average number of interacting atoms within the volume $V=\frac{4\pi}{3}(r_0^3-r_{\mathrm{B}}^3)$ (the higher abscissa). The population of the $42$D$_{5/2}$ is set to $\sim3.3\,\%$ in these simulations. The interaction strength is limited below $75\,$GHz, which is the half width half maximum of the pump excitation, and the peak atom density is set to $\sim1.3\times10^{12}$\,cm$^{-3}$ in these simulations. The red dashed and green dotted lines show the results with the dipole-dipole interaction and the hybrid form of a dipole-dipole and a van der Waals interaction without anisotropies, respectively. The results with the van der Waals interaction without an anisotropy, which has been used in the main text, are displayed by the blue solid lines. The dark-grey solid lines represent the measured Ramsey contrast and the phase-shift, each of which is the average over eight points around $\tau=500$\,ps in Figs.~6b or d. The light-grey shaded area represents one standard deviation of the average over those eight measured values. The black solid line shows the Ramsey contrast $|g(\tau)|$ given by Eq.~(7) in the main text with $\gamma(\tau) = 0$, giving the upper limit of the contrast decay and accordingly the lower limit of the number of atoms $\sim 32$ to reproduce the Ramsey contrast $\sim 0.45$ measured at $\tau$ = 500 ps, irrespective of the potential curves.}
		\label{Suppfig09}
	\end{center}
\end{figure}


\begin{figure}[ht]
	\begin{center}
		\includegraphics[scale=0.82]{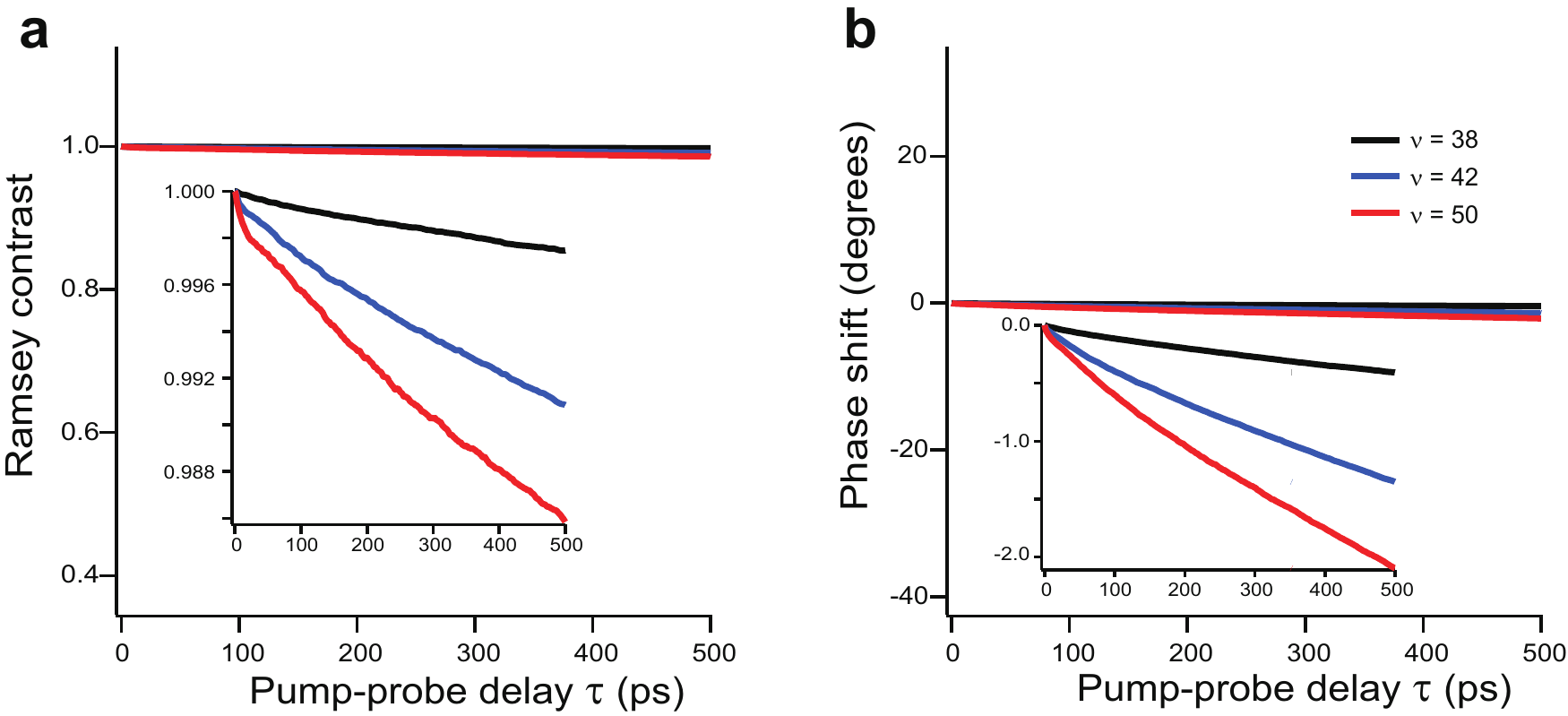}
		\caption{{\bf Numerical simulation of the photo-ions contribution to the Ramsey signal.} (\textbf{a}) Ramsey-contrasts simulated as functions of pump probe delay $\tau$ in the presence of photo-ions for $\nu=38$ (black), $42$ (blue), and $50$ (red). (\textbf{b}) Phase-shifts simulated as functions of pump probe delay $\tau$ in the presence of photo-ions for $\nu=38$ (black), $42$ (blue), and $50$ (red).}
		\label{Suppfig11a}
	\end{center}
\end{figure}


\begin{figure}[t]
	\begin{center}
           \includegraphics[scale=0.8]{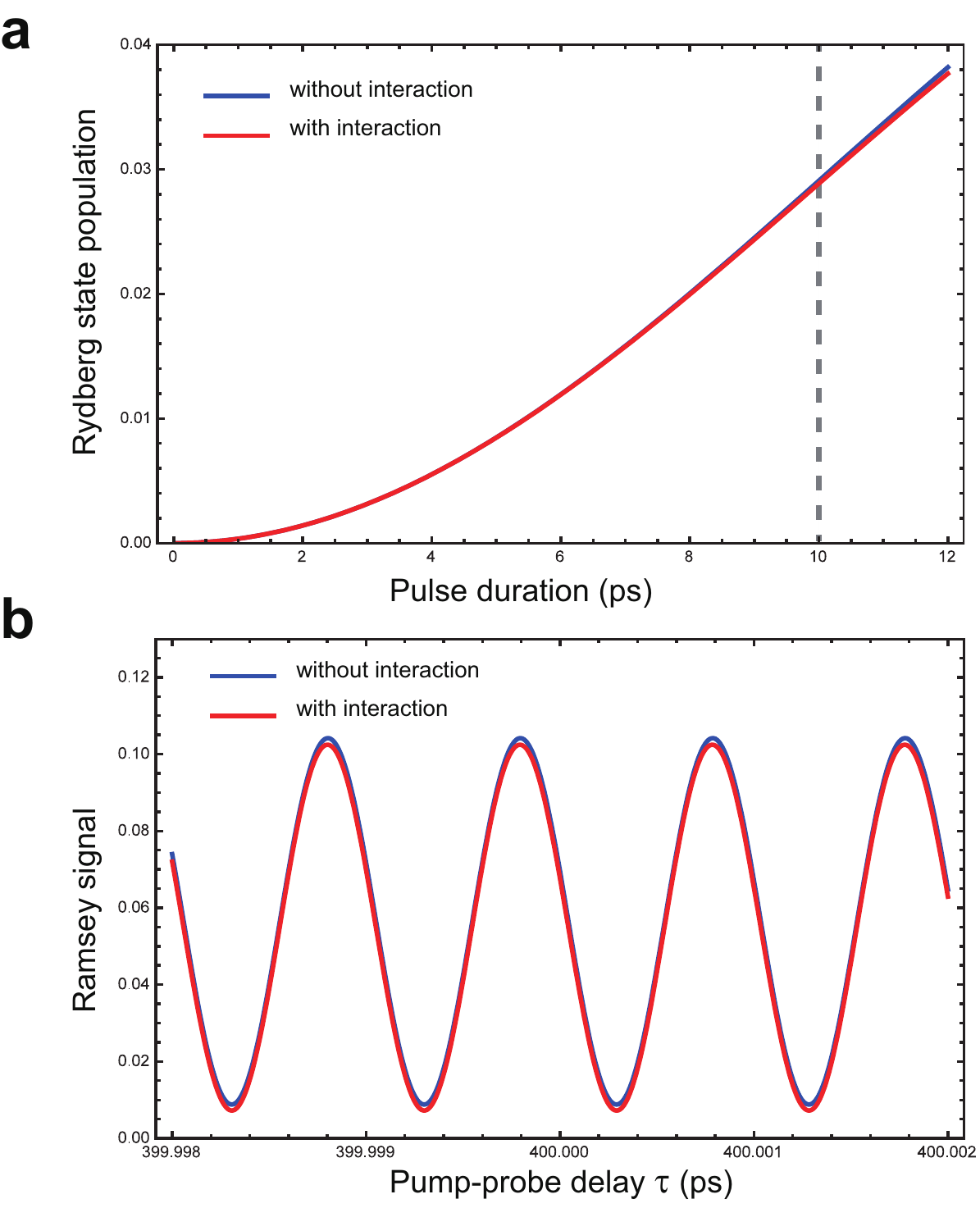}
           \caption{{\bf Simulation of the effects of the Rydberg interactions during the excitation on the Ramsey signal.} (\textbf{a}) The Rydberg populations at the end of the excitation pulse as functions of the pulse duration with (red) and without (blue) the interaction. (\textbf{b}) the Ramsey oscillations around the pump-probe delay $\tau \sim 400$\,ps with (red) and without (blue) the interaction during the excitation pulses.}
		\label{Suppfig02}
	\end{center}
\end{figure}


\begin{figure}[t]
	\begin{center}
		\includegraphics[scale=0.85]{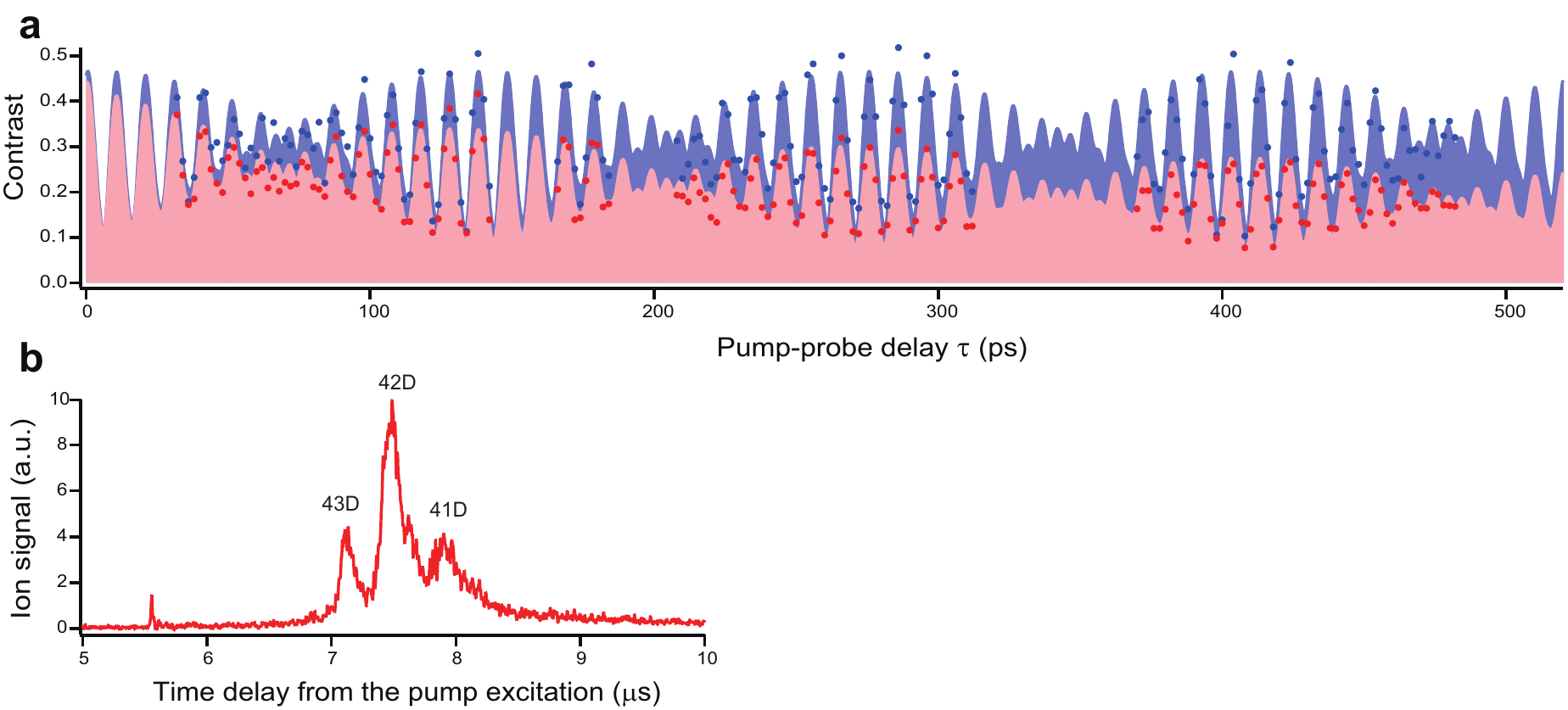}
		\caption{{\bf High resolution measurement of the Ramsey-oscillation contrasts.} (\textbf{a}) The contrasts as functions of the pump-probe delay $\tau$ for the higher-density (red dots) and lower-density (blue dots) ensembles, respectively. The red- and blue-shaded parts are numerically simulated recurrence-motions of the Rydberg wave-packet for the higher- and lower-density ensembles, respectively. (\textbf{b}) Field ionization spectrum associated with the measurement shown in {\bf a}.}
		\label{Suppfig11wp}
	\end{center}
\end{figure}


\begin{figure}[t]
	\begin{center}
		\includegraphics[scale=0.6]{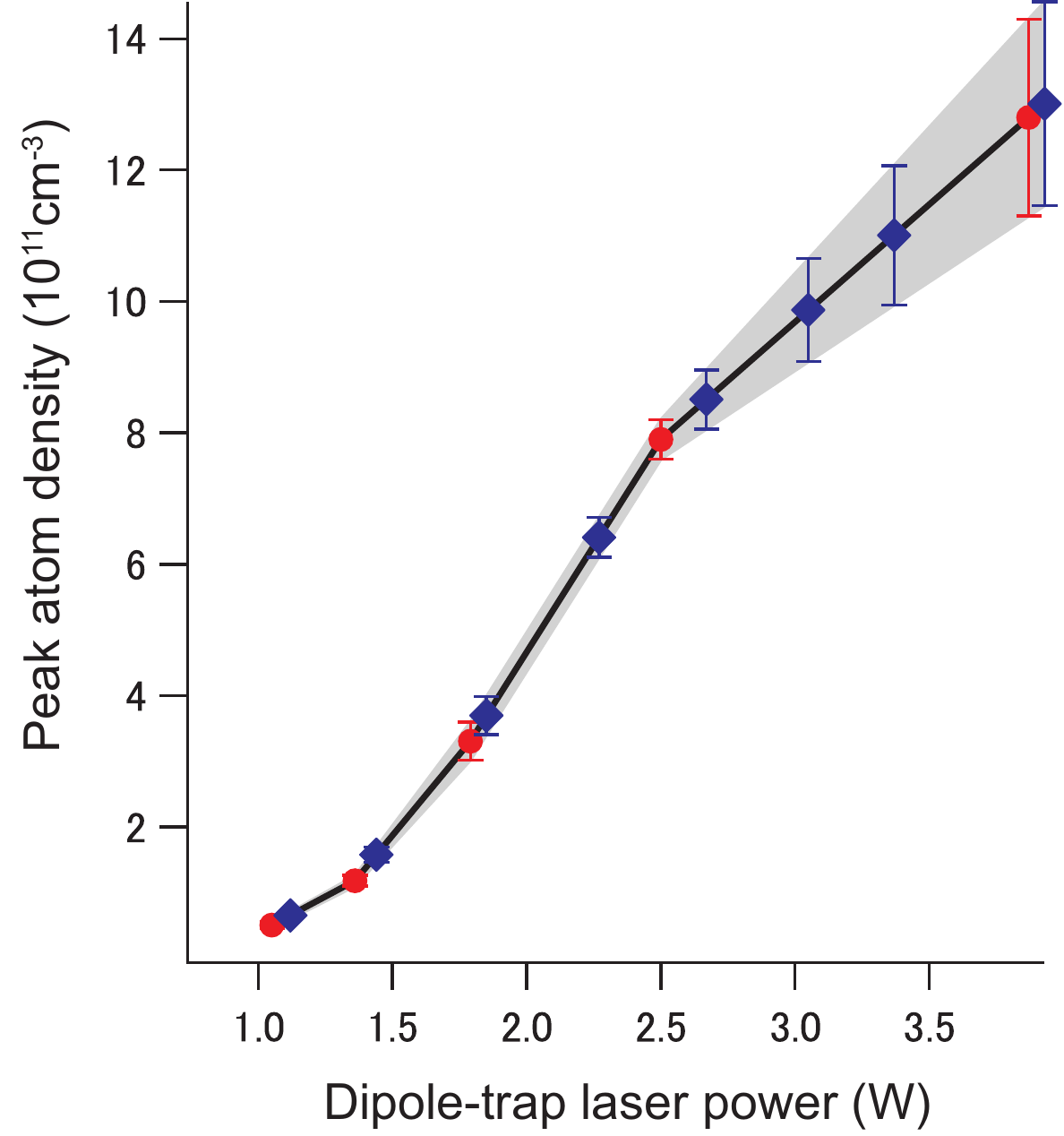}
		\caption{{\bf The calibration of the peak atom density as a function of the power of the dipole-trap-laser.} The red-circle data points show the peak atom density measured in the calibration experiment as a function of the dipole-trap-laser power. The blue-diamond-shaped data and their error bars are not measured directly, but are obtained by linear interpolation. These blue-diamond-shaped data points represent the estimated peak atom densities used in the density dependence measurement of the Ramsey contrast shown in Fig.~4b in the main text. The error bars represent one standard deviation.}
		\label{Suppfig12}
	\end{center}
\end{figure}


\begin{figure}[ht]
	\begin{center}
		\includegraphics[scale=0.67]{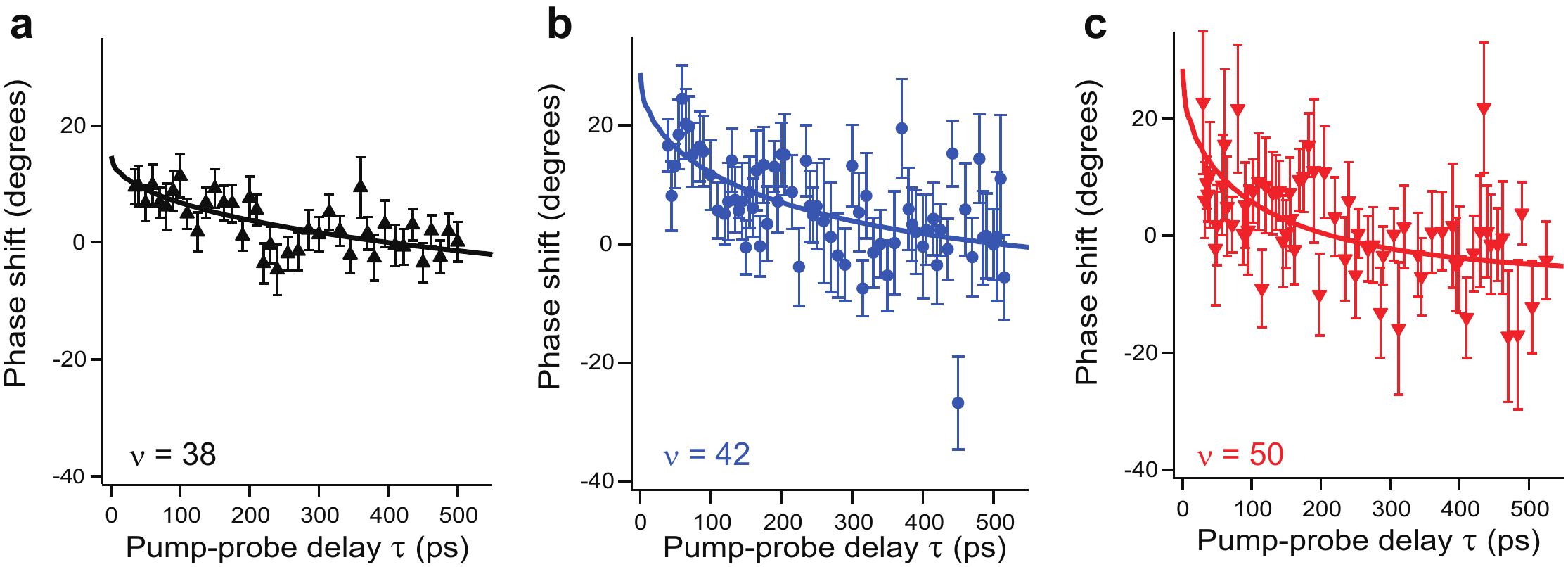}
		\caption{{\bf The principal-quantum-number dependences of the phase-shift.} (\textbf{a} -- \textbf{c}) Measured phase-shifts are plotted as functions of $\tau$ for three different Rydberg levels $\nu=38$, $42$, and $50$. Each of them has been measured simultaneously with the Ramsey contrast with the same principal-quantum-number shown in Fig.~4a in the main text. The simulations indicated by the black, blue, and red solid-lines have been performed by the theory model with the continuum approximation for the van der Waals interaction with the adjusting parameters being $C_6=8$\,GHz\,$\mu$m$^6$ for $\nu=38$, $C_{6} = 34\,$GHz\,$\mu$m$^{6}$ for $\nu=42$, and $C_6=103$\,GHz\,$\mu$m$^6$ for $\nu=50$. The peak atom-density is set to the estimated density for each Rydberg level in these simulations. The density estimations are described in Methods section ``Estimation of the atom density'' in the main text. It should be noted that several Rydberg states are excited in the case of $\nu=50$, whereas we have considered an excitation only to a single Rydberg state to perform the simulations for all of the three Rydberg levels. The error bars represent the standard deviation.}
		\label{Suppfig11}
	\end{center}
\end{figure}


\begin{figure}[ht]
	\begin{center}
		\includegraphics[scale=0.8]{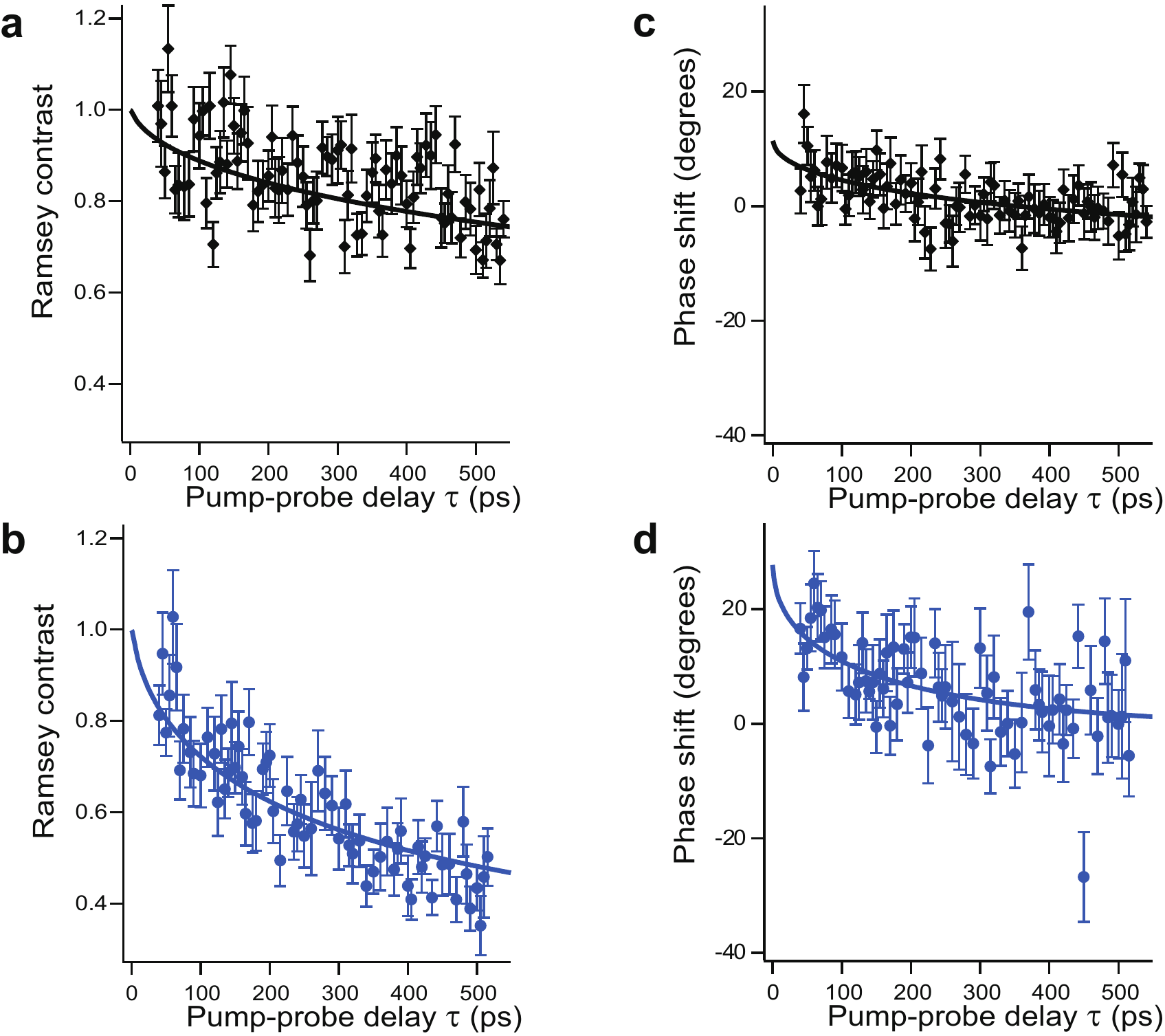}
		\caption{{\bf The theory-model analysis of the Ramsey contrast and phase-shift beyond mean-field with the continuum approximations and with an anisotropic van der Waals potential.} The black-diamond-shaped and blue-circle data-points show the Ramsey contrasts (\textbf{a} and {\bf b}) and the phase-shifts (\textbf{c} and {\bf d}) measured with the population of the $42$D$_{5/2}$ state being $\sim$1.2$\%$ (\textbf{a} and {\bf c}) and $\sim$3.3$\%$ (\textbf{b} and {\bf d}), respectively. In \textbf{a} and \textbf{b}, the Ramsey contrasts simulated with the anisotropic potential given by Supplementary Eq.~(\ref{eq:Aniso1}) (black and blue solid lines) are compared with the measured ones. Similarly, the measured and simulated phase-shifts are compared in \textbf{c} and \textbf{d}. These simulated results have been obtained by Eq.~(6) in the main text combined with the potential anisotropy, employing the cutoff radius $r_{0} \sim 4\,\mu$m and its corresponding atom-number $N_{0} = 450$ at the peak density. The interaction strength in these simulations is limited below $75\,$GHz, which is the half width half maximum of the pump excitation, and the peak atom density is set to $\sim1.3\times10^{12}$\,cm$^{-3}$ in these simulations. The coefficient $C_{6} = 63\,$GHz\,$\mu$m$^{6}$ has been used in these simulations. The error bars represent the standard deviation.}
		\label{Suppfig05}
	\end{center}
\end{figure}


\begin{figure}[H]
	\begin{center}
		\includegraphics[scale=0.8]{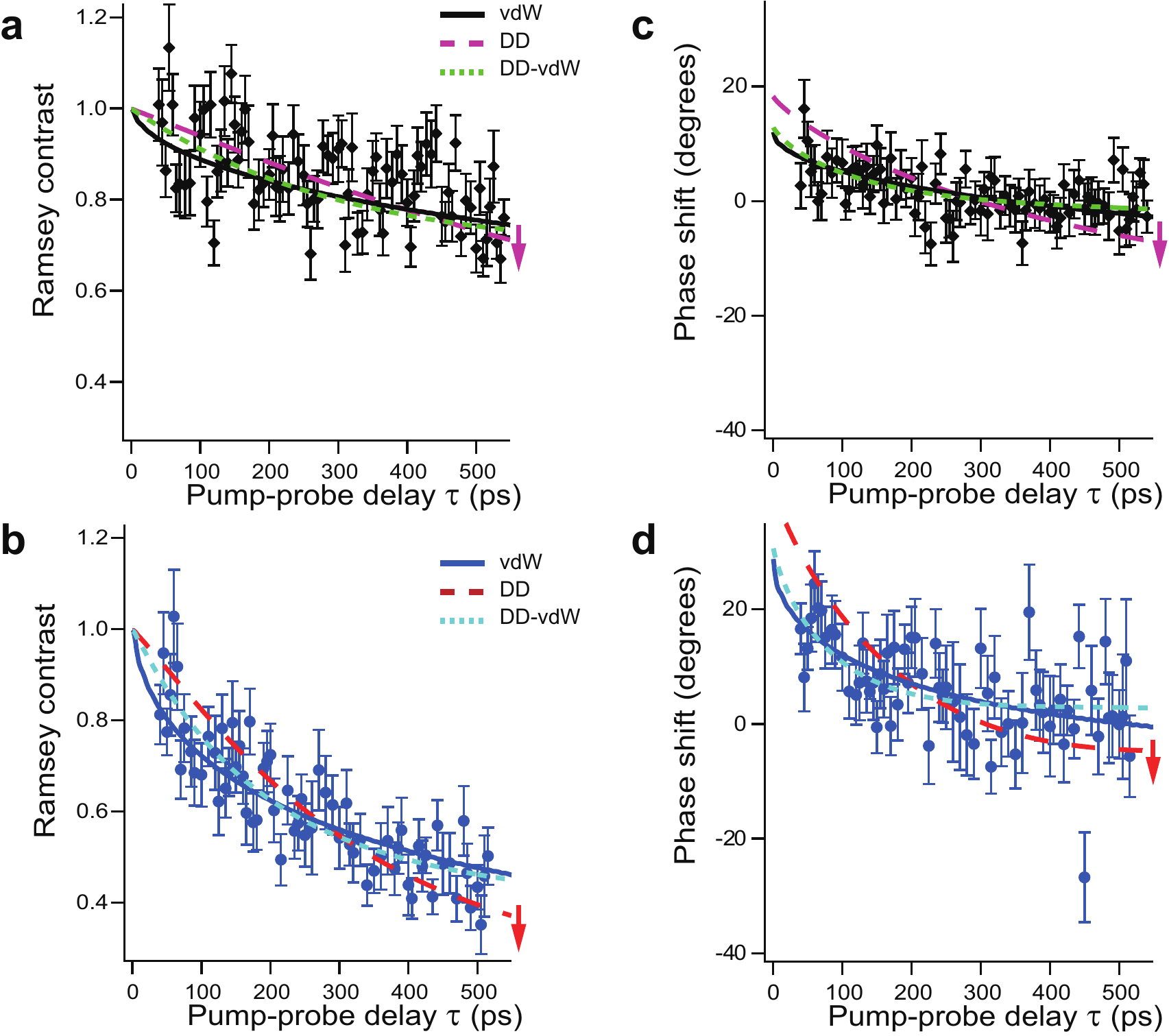}
		\caption{{\bf The theory-model analysis of the Ramsey contrast and phase-shift beyond mean-field with the continuum approximation and with the dipole-dipole (DD) interaction and the hybrid form of a dipole-dipole and a van der Waals (DD-vdW) interaction without anisotropies.} The black-diamond-shaped and blue-circle data-points show the Ramsey contrasts (\textbf{a} and {\bf b}) and the phase-shifts (\textbf{c} and {\bf d}) measured with the population of the $42$D$_{5/2}$ state being $\sim$1.2$\%$ (\textbf{a} and {\bf c}) and $\sim$3.3$\%$ (\textbf{b} and {\bf d}), respectively. In \textbf{a} and \textbf{b}, the Ramsey contrasts simulated with the DD (dahed line) and DD-vdW (dotted line) interactions without anisotropies are compared with the measured ones. Similarly the measured and simulated phase-shifts are compared in \textbf{c} and \textbf{d}. The simulated results for the DD interaction have been obtained by Eq.~(6) in the main text, employing the cutoff radius $r_{0} \sim 4\,\mu$m and its corresponding atom-number $N_{0} = 450$ at the peak density. It should be noted that these DD results are pushed down as the cutoff radius is further increased, as is indicated by arrows in the figures and will be further discussed with Supplementary Fig.~\ref{Suppfig09}. The simulations with the pure van der Waals (vdW) interaction (solid line), which have been shown in Fig.~6 in the main text, are presented again to be compared with the DD and DD-vdW results. The interaction strength is limited below $75\,$GHz, which is the bandwidth (half width half maximum) of the pump excitation, and the peak atom density is set to $\sim1.3\times10^{12}$\,cm$^{-3}$ in these simulations. The coefficient $C_{3} = 2.5\,$GHz\,$\mu$m$^{3}$ have been used for the DD interaction, and a combination of $C_{3} = 4.3\,$GHz\,$\mu$m$^{3}$ and $r_{c} = 1.93\,\mu$m have been used for the DD-vdW interaction~[see Supplementary Eq.~\eqref{eq:Eff1}]. The error bars represent the standard deviation.}
		\label{Suppfig07}
	\end{center}
\end{figure}

\clearpage
\thispagestyle{empty}

\begin{figure}[ht]
	\begin{center}
		\includegraphics[scale=0.8]{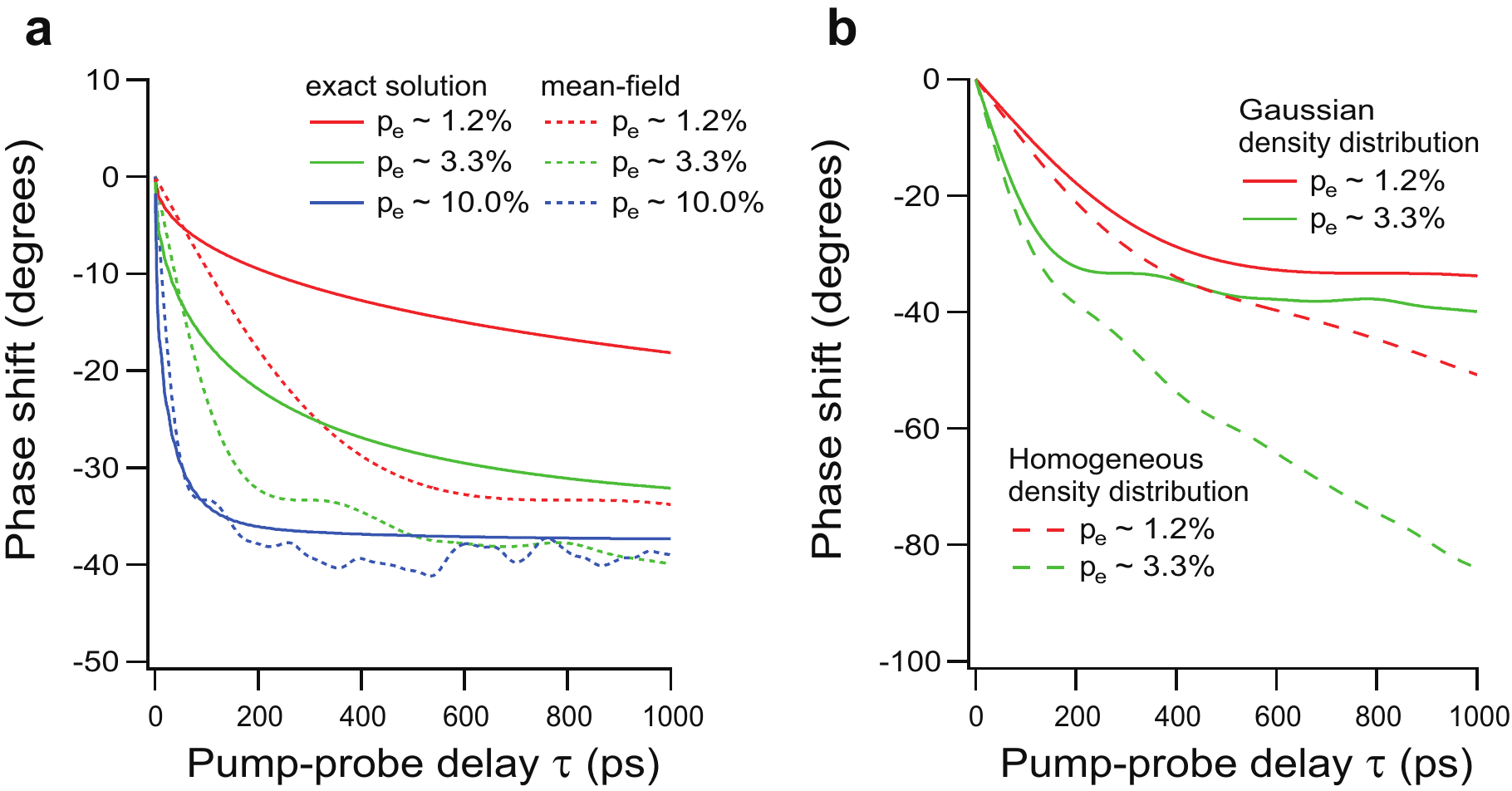}
		\caption{{\bf Phase-shifts calculated as functions of the pump-probe delay $\tau$.} (\textbf{a}) Comparison between the exact (solid lines) and mean-field (dotted lines) solutions with the Gaussian atom-density distribution for different populations $p_\mathrm{e}$ ($\sim1.2$, $\sim3.3$, and $\sim10.0\,\%$) in the 42D Rydberg state. The Gaussian distribution has been taken from the Ramsey experiment shown in the main text. The $C_6$ coefficients are set to $34\,\mathrm{GHz}\,\mu\mathrm{m}^6$ for the exact solution and $1.9\,\mathrm{GHz}\,\mu\mathrm{m}^6$ for the mean-field solution, respectively, which are the values used in the calculations performed in the main text. (\textbf{b}) Comparison between phase-shifts calculated with the Gaussian (solid lines) and homogenous (dashed lines) atom-density distributions within the mean-field model for different populations $p_\mathrm{e} \sim1.2\,\%$ and $\sim3.3\,\%$ in the 42D Rydberg state. The Gaussian distribution is the same as in \textbf{a}, whereas the homogeneous density has been set to the average density of the Gaussian distribution. The $C_6$ value is set the same as in \textbf{a}.}
		\label{Suppfig16}
	\end{center}
\end{figure}


\noindent
{\bf {\Large Supplementary Notes}}\\

\noindent
{\bf {\large Supplementary Note 1: Estimation of the photo-ion contribution to the Ramsey signal}}

In order to investigate the effect of the electric field generated by photo-ions on the Ramsey signal, we have performed numerical simulations for $\nu=38$, $42$, and $50$, including ion contributions. In each of those simulations, atom locations in an ensemble with a Gaussian distribution are generated by the Monte-Carlo method. In this list of atom locations a certain number of random locations are taken to be ion locations. The number of ions is obtained as follows.
The ion fractions are estimated for $\nu=38$ and $50$ from the field ionization spectra shown in Supplementary Figs.~\ref{Suppfig10}a and c, respectively, and for $\nu=42$ from the spectrum similar to the one shown in Supplementary Fig.~\ref{Suppfig10}b to be the ratios of the areas of the ion peaks at $\sim 5.6\,\mu$s for $\nu=38$, $42$ and $\sim 6.2\,\mu$s for $\nu=50$ to the total areas of the spectra.
The numbers of ions are thus estimated as the products of those ion fractions, our Rydberg populations $p_\mathrm{e}$'s, and the total number of atoms to be $109$ for $\nu=38$ with $p_\mathrm{e} \sim 3.2\,\%$, $232\,$ for $\nu=42$ with $p_\mathrm{e} \sim 3.3\,\%$, and $108$ for $\nu = 50$ with $p_\mathrm{e} \sim 3.1\,\%$, respectively. These numbers are the upper limits of the number of photo-ions as they may include the ions produced at much later time ($100\,$ns timescale) by Rydberg interactions $[1]$. At each atom position in each of the ensembles with these ion numbers, the joint electric field generated by those ions is calculated. The Stark shift induced by this joint electric field is calculated for each atom, so that the period of its Ramsey oscillation is shifted accordingly. Those Ramsey oscillations are averaged over the whole ensemble of the Rydberg atoms to give the contrast-decay and phase-shift shown in Supplementary Fig.~\ref{Suppfig11a} for each of $\nu=38$, $42$, and $50$. It is seen that the contrast-decay and phase-shift are less than $2\,\%$ and $2.5$ degrees at the longest pump-probe delay $500$\,ps in our measurements and are almost negligible.\\


\noindent
{\bf {\large Supplementary Note 2: Discussion on the pump-probe excitation and the zero-delay offset in the phase-shift}}

We consider the time-domain Ramsey interferometry with the identical pump and probe excitations, each of which is a one-photon excitation of a two-level atom, consisting of a ground state $|\mathrm{g}\rangle$ and an excited state $|\mathrm{e}\rangle$, with rectangular laser pulses for simplicity. The results can be easily adapted to near-resonant two-photon excitations that have been used in the present experiments.

The time dependent Schr\"odinger equation is given by
\begin{eqnarray*}
\label{eq:text0}
i\hbar \frac{\partial}{\partial t} |\psi \rangle &=& H|\psi \rangle \\\nonumber
&=& \left( \hbar \frac{\omega}{2} |\mathrm{e}\rangle\langle \mathrm{e}| - \hbar \frac{\omega}{2} |\mathrm{g}\rangle\langle \mathrm{g}| -{\bf d}_\mathrm{eg}\cdot{\bf E}(t)|\mathrm{e}\rangle\langle \mathrm{g}|  -{\bf d}_\mathrm{eg}^{*}\cdot{\bf E}(t)|\mathrm{g}\rangle\langle \mathrm{e}|\right) |\psi \rangle.
\end{eqnarray*}
Here, $\omega$ is the angular frequency for the atomic transition, ${\bf E}(t) = {\bf E}_{0}(\Theta(t-t_{0})-\Theta(t-t_{1}))\cos(\omega_{l}(t-t_{0}))$ is the pump laser field where $\omega_\mathrm{l}$ is the laser frequency, $\Theta(t)$ is the Heaviside step function, $t_{0}$ is the beginning of the pulse excitation, $t_{1}$ is the end of the excitation, $|\psi\rangle = C_\mathrm{g}(t)|\mathrm{g}\rangle + C_\mathrm{e}(t)|\mathrm{e}\rangle$ is the atomic state, and ${\bf d}_\mathrm{eg} = \langle \mathrm{e}| {\bf d}|\mathrm{g}\rangle$ is the dipole-matrix element.
Solving the Schr\"odinger equation Eq.~(\ref{eq:text0}) by employing the rotating-wave approximation for the evolution of the atomic state during the pump excitation we obtain 
\begin{eqnarray}
\label{eq:text1}
\nonumber
\left(\begin{array}{c} C_\mathrm{g}(t_{1})\\  C_\mathrm{e}(t_{1}) \end{array}\right) &=& \left ( \begin{array}{cc} (\cos(\theta/2)-i\eta\sin(\theta/2))e^{i\frac{\omega_\mathrm{l}}{2}\delta t} & -i\xi\sin(\theta/2))e^{i\chi}e^{i\frac{\omega_\mathrm{l}}{2}\delta t} \\ -i\xi\sin(\theta/2)e^{-i\chi}e^{-i\frac{\omega_\mathrm{l}}{2}\delta t} & (\cos(\theta/2)+i\eta\sin(\theta/2))e^{-i\frac{\omega_\mathrm{l}}{2}\delta t} \end{array} \right ) \left(\begin{array}{c} C_\mathrm{g}(t_{0})\\ C_\mathrm{e}(t_{0}) \end{array}\right)\\
&=& A \left(\begin{array}{c} C_\mathrm{g}(t_{0})\\ C_\mathrm{e}(t_{0}) \end{array}\right),
\end{eqnarray}
with $\theta = \Omega \delta t$, $\eta = \Delta/\Omega$, $\xi = \Omega_\mathrm{R}/\Omega$, and $\Omega = \sqrt{\Omega^{2}_\mathrm{R} + \Delta^{2}}$, where $\Omega_\mathrm{R}$ is the Rabi frequency, and $\Delta = \omega -\omega_\mathrm{l}$ is the detuning of the laser frequency $\omega_\mathrm{l}$ from the atomic resonance $\omega$. The time $\delta t=t_{1}-t_{0}$ is the duration of the pump excitation, and $\chi$ represents the phase of the complex Rabi frequency $\tilde{\Omega}_\mathrm{R} = \Omega_\mathrm{R}e^{i\chi}$. It is important to note in the formulation given in Eq.~(\ref{eq:text1}) that the matrix $A$, which describes the time evolution of the atomic state during the pump excitation, is given in the lab frame, in which the field-free evolution of the atomic state between the pump and probe excitations is described.

The matrix for the excitation can be expressed in the form
\begin{eqnarray}
\label{eq:text2}
A = \left ( \begin{array}{cc} |c_\mathrm{g}|e^{i\phi/2} & i|c_\mathrm{e}|e^{i\zeta/2} \\ i|c_\mathrm{e}|e^{-i\zeta/2} & |c_\mathrm{g}|e^{-i\phi/2}
\end{array} \right ),
\end{eqnarray} 
where $\phi$ and $\zeta$ are phases of the complex amplitudes $c_\mathrm{g}$ and $c_\mathrm{e}$, respectively.
The pump and probe excitation operation on the ground state $|\mathrm{g}\rangle$ results in
\begin{eqnarray}
\label{eq:text2a}
|\psi(\tau) \rangle &=& A\exp(-iH\tau/\hbar)A |\mathrm{g} \rangle \\\nonumber
&=& \left( \begin{array}{c} |c_\mathrm{g}|^{2}e^{i\phi}e^{i\frac{\omega}{2}\tau} - |c_\mathrm{e}|^{2}e^{-i\frac{\omega}{2}\tau} \\  i|c_\mathrm{g}||c_\mathrm{e}|e^{-i\zeta/2}(e^{i\phi/2}e^{i\frac{\omega}{2}\tau} + e^{-i\phi/2}e^{-i\frac{\omega}{2}\tau}) \end{array}\right)
\end{eqnarray}
Therefore, the population in the excited state $P(\tau) =\langle \psi(\tau)|\mathrm{e}\rangle\langle \mathrm{e}|\psi(\tau) \rangle$ that remains after the probe excitation is given by
\begin{eqnarray}
\label{eq:App3}
P(\tau) &=& 2|c_\mathrm{g}|^{2}|c_\mathrm{e}|^{2}\left(1 + \cos(\omega\tau+\phi)\right).
\end{eqnarray}
Here only the  phase $\phi$ remains in the Ramsey signal, while the phase $\zeta$ is cancelled out. According to Eq.~(\ref{eq:text1}) and Eq.~(\ref{eq:text2}) the phase $\phi$ is defined by
\begin{eqnarray}
\label{eq:text3} 
\tan(\phi/2) &=& \frac{\Omega\sin(\omega_\mathrm{l}\delta t/2)\cos(\Omega\delta t/2)-\Delta\cos(\omega_\mathrm{l}\delta t/2)\sin(\Omega\delta t/2)}{\Omega \cos(\omega_\mathrm{l}\delta t/2)\cos(\Omega\delta t/2)+\Delta\sin(\omega_\mathrm{l}\delta t/2)\sin(\Omega\delta t/2)}.
\end{eqnarray} 
When the laser detuning vanishes ($\Delta = 0$) we obtain $\phi = \omega_\mathrm{l}\delta t = \omega \delta t$. Similarly, if $\Omega_\mathrm{R} \to 0$, the phase $\phi$ converges to $\omega \delta t$.
In the regime with $|\Delta| >> |\Omega_\mathrm{R}|$ where $\Omega = \sqrt{\Omega^{2}_\mathrm{R} + \Delta^{2}} \approx \Delta + \frac{\Omega_\mathrm{R}^{2}}{2\Delta}$, the effects of the laser field can be characterized by the AC-Stark shift $\omega_\mathrm{ac} = \frac{\Omega_\mathrm{R}^{2}}{4\Delta}$, so that we obtain  
\begin{eqnarray}
\label{eq:text4} 
\tan(\phi/2) &\approx& \tan\left(\left(\omega - \frac{\Omega_\mathrm{R}^{2}}{2\Delta}\right)\delta t/2 \right).
\end{eqnarray} 
Therefore the phase aquired during the excitation is given by the AC-Stark-shifted phase-evolution 
\begin{eqnarray}
\label{eq:App9}
\phi \approx \left( \omega - 2\omega_\mathrm{ac}\right) \delta t.
\end{eqnarray}
If the interactions among the atoms are neglected during the ps pulses, this intrapulse phase is common to the higher- and lower-density ensembles under the condition that their sizes, shapes, and positions are identical. Accordingly, the intrapulse phase is cancelled out in the phase-shift between those two ensembles. 

In the actual measurements, however, slight differences between the sizes, shapes, and positions of the higher- and lower-density ensembles can lead to different intrapulse phases due to different AC-Stark shifts (see Methods section ``Estimation of the atom density'' in the main text). Additional intrapulse phases due to the Rydberg interactions during the pump and probe excitations are considered to be negligibly small as seen in the theoretical simulations demonstrated in the next paragraph.

We have performed theoretical simulations of the Ramsey signals to investigate the effects of the Rydberg interactions during the pump (or probe) excitation, whose width is $10\,$ps, on the contrast and phase of the Ramsey oscillations. We consider a two-level system with its level spacing that corresponds to a frequency of $1.008\times10^{15}\,$Hz and is the energy difference between 5S and 42D states. We assume the van der Waals interaction $C_6 \sim$ 34 GHz\,$\mu$m${}^6$ as it has been obtained in the beyond-mean-field analyses of our data in the main text. The interatomic distance that gives the interaction timescale comparable to the width of the excitation $\sim10\,$ps is estimated to be  $\sim 0.84\,\mu$m. The average number of atoms within a sphere whose radius is $0.84\,\mu$m is estimated to be less than two in our higher-density ensemble. The number of atoms $N$ that has been considered in the present simulation is five and is thus large enough to investigate the interaction effects within the  $\sim10\,$ps pulse excitation. We have compared the populations of the excited state generated by an effective off-resonant two-photon excitation with and without the atom-atom interactions during the excitation process. We have considered the atoms to be randomly distributed within a cube with a volume $r^3_0 = N/n_\mathrm{av}$ where the average atom density $n_\mathrm{av} = (1.3/2^{(3/2)})\times10^{12}\,$cm$^{−3}$ is the same value as the one in our experiment.
We plot the excited state populations at the end of the excitation pulse as functions of the pulse duration with and without the atom-atom interactions in Supplementary Fig.~\ref{Suppfig02}a.
Here the red and blue curves show the results with and without the interactions, respectively, during the excitation. We fixed the average Rabi frequency to $18.88\,$GHz, and the detuning from resonance emerging due to the two-photon light shift to 150 GHz, so that the resulting populations are $\sim3\,\%$ in the excited state at the pulse duration of  $10\,$ps. It is seen in Supplementary Fig.~\ref{Suppfig02}a that the populations are almost the same between these two cases with their population difference being less than $1\,\%$ at $10\,$ps, indicating that the interaction effects on the excited-state population are almost negligible during the pulse excitation. Next we have simulated the Ramsey signals with the pump and probe excitations and evaluated the intrapulse-interaction effects on the contrast and phase of the Ramsey oscillations. The simulated Ramsey oscillations are shown in Supplementary Fig.~\ref{Suppfig02}b where the red and blue curves show the results with and without the interactions during the excitation pulses, respectively, around the pump-probe delays $\tau=400\,$ps. We calculate the contrasts for those traces and obtain $0.867$ and $0.844$ with and without the interactions, respectively. The difference between the contrasts is thus $\sim3\,\%$, originating from the interaction during the the excitations. The phase difference between those two traces is less than $\sim 1\,$degrees. This corrsponds to the intrapulse phase discussed in the preceding paragraph. These simulated values for the contrast reduction and the phase difference are much smaller that the measured contrast-decay and the amount of phase-shift seen in Figs.~3 and 6. It is thus concluded that the interaction effects on the contrast and phase are almost negligible. It is also concluded that the contribution of the interactions to the intrapulse delay is negligibly small.

We have verified that the discussion above holds entirely also for a three-level system with rectangular laser pulses. We consider a system consisting of a ground state $|\mathrm{g}\rangle$, an intermediate state $|\mathrm{i}\rangle$, and a Rydberg excited state $|\mathrm{e}\rangle$. A laser field with frequency $\omega_\mathrm{l1}$ couples the state $|\mathrm{g}\rangle$ to the upper state $|\mathrm{i}\rangle$ with Rabi frequency $\Omega_\mathrm{R1}$ and detuning $\Delta_1\equiv \omega_\mathrm{l1}-\omega_1$, where $\omega_1$ is the corresponding transition frequency. A second laser field with Rabi frequency $\Omega_\mathrm{R2}$ drives the transition between the states $|\mathrm{i}\rangle$ and $|\mathrm{e}\rangle$ with the transition frequency $\omega_{2}$. In the case that $|\Delta_{1}| \gg |\Omega_\mathrm{R1}|, |\Omega_\mathrm{R2}|$, we can neglect the excitation to the intermediate state,  so that the three-level system is effectively reduced to a two-level system. 
In our actual experiments, however, we use an excitation laser-pulse whose envelop is not rectangular. The condition $|\Delta_{1}| \gg |\Omega_\mathrm{R1}|, |\Omega_\mathrm{R2}|$ holds in the beginning and end of the pulse, whereas it might be violated at its peak intensity around the middle of the pulse, so that the intermediate state $|\mathrm{i}\rangle$ can be populated at the peak intensity. However, this state $|\mathrm{i}\rangle$, which is not a Rydberg state, has a negligibly small intrapulse-interaction effect during the excitation pulse and due to adiabaticity given by the condition $\omega_{1}, (\omega_{2}-\omega_{1}) \gg |\Omega_\mathrm{R1}|, |\Omega_\mathrm{R2}|$ no population remains in the intermediate state $|\mathrm{i}\rangle$ after the pulse. This shows that our two-photon excitation can be effectively treated as a one-photon excitation. The discussions above on the intrapulse phase and the interaction effects during the excitation also apply to the two-photon excitation in our current experiments.\\

\noindent
{\bf {\large Supplementary Note 3: Field ionization spectra and estimation of excitation bandwidth}}

In order to reduce the number of Rydberg states to be excited, the spectra of the ps IR and blue pulsed lasers were cut to be about 0.13 nm and 0.20 nm, respectively, with pulse shapers in a 4f configuration ($f=500$\,mm). These bandwidths were common to the Ramsey measurements of the three different Rydberg states $\nu=38$, $42$, and $50$. Supplementary Figure~\ref{Suppfig10} shows state-resolved field-ionization spectra measured by ramping the electric field slowly on the microsecond timescale. The spectra indicate that a single state was predominantly populated for each of the excitations to $\nu=38$ and $42$ as seen in Supplementary Figs.~\ref{Suppfig10}a and b, whereas more levels were populated for the excitation to $\nu=50$ as seen in Supplementary Fig.~\ref{Suppfig10}c. This is because the energy levels of Rydberg states are inversely dependent on $\nu^2$ and more congested for higher states. Due to the characteristics of field ionization, the threshold value of the electric field to induce ionization depends on $\nu^{-4}$~$[2]$, so that higher states are less resolvable with the same ramp-up speed of the electric field. Therefore the ramp-up speed was slower for $\nu=50$ than for $\nu=38$ and $42$ by a factor of $2/3$.

The bandwidth of our Rydberg excitation with the IR and blue pulses was determined from the field-ionization spectrum for $\nu=42$ presented in Fig.~1c as well as in Supplementary Figs.~\ref{Suppfig10}b.
Assuming a symmetrical population distribution with respect to the center level $\nu=42$, we estimated the relative populations in the excited Rydberg states to be $\sim84\,\%$ for the center state 42D and $\sim8\,\%$ for each neighboring state 43D and 41D, respectively, from the integrated areas of the ion signal peaks. In Supplementary Figs.~\ref{Suppfig10}b, the integration range for the 43D state (the 42D and 41D states) is indicated by the red-shaded (blue-shaded) region. We did not deconvolute the ion spectrum for those integrations although those peaks were not fully resolved.
Based on the relative populations and assuming a Gaussian excitation spectrum, we obtained a bandwidth of $\sim$150\,GHz (FWHM).\\

\noindent
{\bf {\large Supplementary Note 4: Structures in the pump-probe delay dependence of the Ramsey signal}}

We have made another set of supplementary Ramsey measurements with smaller steps of the pump-probe delay $\tau$ than those of the measurements shown in the main text. The results of those supplementary measurements are shown in Supplementary Fig.~\ref{Suppfig11wp}a, in which the red and blue dots show the contrasts of the Ramsey oscillation for the higher- and lower-density ensembles, respectively, as functions of $\tau$. It should be noted here that Supplementary Fig.~\ref{Suppfig11wp}b shows a field ionization spectrum associated with the Ramsey measurement shown in Supplementary Fig.~\ref{Suppfig11wp}a, indicating the 41D, 42D, and 43D states predominantly populated, similar to the measurements shown in the main text, but more fractions of the 41D and 43D states included in these supplementary measurements with a broader bandwidth of the ps excitation laser pulse. It is thus reasonable that the recurrence motion of the Rydberg wave-packet with a period of $\sim10$\,ps is more pronounced and better resolved in Supplementary Fig.~\ref{Suppfig11wp}a than in Fig.~2d in the main text due to the more fractions of the neighboring levels and smaller steps of $\tau$, respectively. The red- and blue-shaded parts shown in Supplementary Fig.~\ref{Suppfig11wp}a correspond to the recurrence motions of the Rydberg wave-packet numerically simulated only with those three Rydberg levels and with the decay factor exp(-$\alpha \sqrt{\tau}$) seen in Eq.~(5) in the main text. It is seen from this figure that the simulated results show excellent agreements with the measured $\tau$ dependences of the contrast, demonstrating that our Ramsey signal is not affected by other angular momentum states such as S, P, and F states or by the oscillation resulting from the ground-state hyperfine splitting.

It is understood from the comparison between the structures seen at $\tau \sim130-170\,$ps in Fig.~2d in the main text and Supplementary Fig.~\ref{Suppfig11wp}a that those oscillations can be assigned to the recurrence motion of the wave-packet. It is also understood from the comparison between the collapse and revival of the wave packet seen in Supplementary Fig.~\ref{Suppfig11wp}a, which is due to an anharmonicity of the Rydberg levels, and the global structure seen in Fig.~2d that the structures on the $\sim 100$\, ps timescale seen in Fig.~2d are not always assigned to the collapse and revival, but also to experimental fluctuations.\\

\noindent
{\bf {\large Supplementary Note 5: Active control of many-body dynamics}}\\
\noindent
{\bf {\large Supplementary Note 5-1: Atom-density dependence of the Ramsey-contrast decay}}

Figure~4b in the main text shows the Ramsey contrasts for $\nu=42$ with a population of $3.5\pm0.3$\,\% at $\tau=300$ and $510$\,ps for several different atom densities ranging from the lower to the higher densities described in the main text. These measurements were made by changing the power of the dipole-trap laser and thereby the trap depth. At first the atom densities in these Ramsey measurements were estimated solely from the total number of atoms and the size of the atomic ensemble obtained by in-situ absorption imaging with a CCD camera without expanding the atomic ensemble, but were underestimated because of the spatial resolution, as is described in Methods section ``Estimation of the atom density'' in the main text. Therefore it was calibrated in a later independent experiment in which the trapping conditions (1) and (3) (the loading sequence and the dipole-trap laser focusing), which were described in Methods section ``Estimation of the atom density'', were almost the same as those employed in the Ramsey measurements. In this calibration experiment, we measured the radial trap-frequency, the temperature, the axial size, and the total number of atoms to obtain the atom density as a function of the power of the dipole-trap laser. Supplementary Figure~\ref{Suppfig12} shows the results of this calibration measurement accompanied by a calibration curve, which is a linear interpolation function. The atom densities in the Ramsey measurements were estimated from this calibration curve as a function of the dipole-trap-laser power as shown by the blue diamond-shaped data points in Supplementary Fig.~\ref{Suppfig12}, and then the Ramsey contrasts measured at $\tau=$300 and 510\,ps are plotted against these estimated atom densities in Fig.~4b in the main text. It is seen in Fig.~4b that the contrast decay is accelerated as the atom density is increased.\\

\noindent
{\bf {\large Supplementary Note 5-2: The principal-quantum-number dependences of the phase-shift}}

Figure~4a in the main text shows the Ramsey contrasts as functions of $\tau$ for three different Rydberg levels $\nu=38$, 42, and 50. The populations $p_\mathrm{e}$ and estimated peak atom-densities are $p_\mathrm{e}\sim3.2\,\%$ and $\sim1.2\times10^{12}\,$cm$^{-3}$ for $\nu=38$, $p_\mathrm{e}\sim3.3\,\%$ and $\sim1.3\times10^{12}\,$cm$^{-3}$ for $\nu=42$, and $p_\mathrm{e}\sim3.1\,\%$ and $\sim1.2\times10^{12}\,$cm$^{-3}$ for $\nu=50$, respectively (see Methods section ``Estimation of the atom density'' in the main text for these density estimations).
It is seen from this figure that the dephasing is accelerated by increasing the principal quantum number $\nu$ of the Rydberg level.
We have also measured the corresponding phase-shifts for these three levels simultaneously with the Ramsey contrasts as shown in Supplementary Fig.~\ref{Suppfig11}.
The theory-model simulations with the continuum approximation indicated by solid lines agree well with the measured results.
It should be noted that several Rydberg states are excited in the case of $\nu=50$, whereas we have considered an excitation only to a single Rydberg state to perform the simulations for all of the three Rydberg levels.\\


\noindent
{\bf {\large Supplementary Note 6: Calculation of the Ramsey contrast and phase-shift with nearest-neighbor interactions}}

We follow previous Ramsey studies on Rydberg interactions~$[3, 4]$ to calculate the Ramsey-contrast-decays and the phase-shifts expected for nearest-neighbor interactions.
In the Ramsey measurement in the main text, the population in the Rydberg state is observed as a function of the delay time $\tau$ between the pump and probe excitations. Within the delay time, nearest neighbor atoms evolve under a Hamiltonian
\begin{eqnarray}
\label{eq:nniH}
H &= & \sum_{j=1}^{2} \left( E_\mathrm{g} |\mathrm{g}\rangle_j \langle \mathrm{g}|_j + E_\mathrm{e} |\mathrm{e}\rangle_j \langle \mathrm{e}|_j\right) +U(r) |\mathrm{e}\rangle_1 |\mathrm{e}\rangle_2 \langle \mathrm{e}|_2 \langle \mathrm{e}|_{1},
\end{eqnarray}
where $|\mathrm{g}\rangle$ and $|\mathrm{e}\rangle$ are ground- and excited Rydberg-states with energies $E_\mathrm{g}$ and $E_\mathrm{e}$, respectively, and $U(r)$ is an interaction energy between a pair of Rydberg atoms separated by $r$.
The time-domain Ramsey signal is obtained by solving a Schr\"odinger equation with this Hamiltonian as
\begin{eqnarray}
\label{eq:nni1}
P(\tau) &= & 2p_\mathrm{g}p_\mathrm{e}\left( 1 + p_\mathrm{g}\cos(\omega\tau+\phi) + p_\mathrm{e}\cos((\omega + U(r)/\hbar)\tau+\phi) \right ) \\\nonumber
&=& 2p_\mathrm{g}p_\mathrm{e}\Re \left\{1 + e^{i(\omega\tau+\phi)} \left(p_\mathrm{g} + p_\mathrm{e}e^{\frac{iU(r)\tau}{\hbar}} \right) \right\}.
\end{eqnarray}
Here $p_\mathrm{g}$ and $p_\mathrm{e}$ are the ground- and Rydberg-state populations, respectively, $\omega=(E_\mathrm{e}-E_\mathrm{g})/\hbar$ is the atomic-resonance frequency, and $\phi$ is the phase offset arising from the AC-Stark shifts during the pulse excitation.
This result is identical to the Ramsey signal obtained by Eq. (3) in the main text with $N=2$.
In a homogeneous atom distribution with a density of $n$, the nearest-neighbor distribution is given by
\begin{eqnarray}
\label{eq:nni2}
Pr(n,r) &=& \exp \left(-\frac{4\pi n r^3}{3} \right) 4\pi n r^2,
\end{eqnarray}
where $\int_0^{\infty}Pr(n,r)dr=1$. The Ramsey signal given by Supplementary Eq.~(\ref{eq:nni1}) is averaged over this distribution as follows:
\begin{eqnarray}
\label{eq:nni3}
P_\mathrm{av}(n,\tau) &=& \int_0^{\infty} Pr(n,r) P(\tau) dr \\\nonumber
&=& 2p_\mathrm{g}p_\mathrm{e}\Re \left\{1 + e^{i(\omega\tau+\phi)} \left(p_\mathrm{g} + p_\mathrm{e} \left[\int_0^{r_\mathrm{B}} Pr(n,r)dr+\int_{r_\mathrm{B}}^{\infty} Pr(n,r) e^{\frac{iU(r)\tau}{\hbar}} dr\right]\right) \right\},
\end{eqnarray}
where $r_\mathrm{B}$ is the blockade radius determined by the finite bandwidth of the excitation with the IR and blue pulses.

This is further averaged over the density distribution of our atomic ensemble, which can be modeled by a Gaussian distribution to be
\begin{eqnarray}
\label{eq:nni4}
n({\bf x}) = n_\mathrm{p}e^{-\frac{(x^{2}+y^{2})}{2\sigma_\mathrm{xy}^{2}}-\frac{z^{2}}{2\sigma_\mathrm{z}^{2}}},
\end{eqnarray}
where $n_\mathrm{p} = \frac{N}{(2\pi)^{3/2}\sigma_\mathrm{xy}^{2}\sigma_\mathrm{z}}$ is the peak density, $N$ is the number of atoms in the ensemble, $\sigma_\mathrm{xy}$ is the width in the x,y-direction, and $\sigma_\mathrm{z}$ is the width in the z-direction of the atomic ensemble. The Ramsey signal thus averaged over the whole ensemble is given by
\begin{eqnarray}
\label{eq:nni5}
P_\mathrm{av}(\tau) &=& \frac{4\pi}{N} \int_{0}^{\infty} d\rho \rho^{2} n(\rho) P_\mathrm{av}(n(\rho),\tau) \\\nonumber
&=& \frac{2}{\sqrt{\pi}n_\mathrm{p}}\int_0^{n_\mathrm{p}} dn \sqrt{\ln\left(\frac{n_\mathrm{p}}{n}\right)} P_\mathrm{av}(n,\tau),
\end{eqnarray}
where the radius $\rho$ is defined by $\rho = \sqrt{ x^{2} + y^{2} + \left(z\frac{\sigma_\mathrm{xy}}{\sigma_\mathrm{z}} \right)^{2}}$. In going from the first line to the second line in Supplementary Eq.~(\ref{eq:nni5}), an integral over the volume of the ensemble is converted to an integral over the density.

It should be noted that the maximum threshold of the Ramsey-contrast-decay expected for nearest-neighbor interactions is determined by the population $p_\mathrm{e}$ of a Rydberg state as follows.
When the Ramsey oscillations are maximally dephased, the third term in the parentheses in the first line of Supplementary Eq.~(\ref{eq:nni1}) vanishes. In this limit, the oscillation contrast and the phase-shift converge to $p_\mathrm{g}\,\,(=1-p_\mathrm{e})$ and $\phi$, respectively, seen in the second term.
These features do not depend on the character of the interaction $U(r)$ such as van der Waals and dipole-dipole.\\

\noindent
{\bf {\large Supplementary Note 7: Mean-field model}}

For the mean-field approximation we follow the notation $|\mathrm{e}\rangle_{j}\langle \mathrm{e}|_{j} = P_\mathrm{ee}^{(j)}$ which results in the following $N$-atom Hamiltonian
\begin{eqnarray}
\label{eq:mean1}
H = \sum_{j=1}^{N} E_\mathrm{e}P_\mathrm{ee}^{(j)}+ \sum_{j=1}^{N-1}\sum_{i>j}^{N} U(r_{i,j})P_\mathrm{ee}^{(i)}P_\mathrm{ee}^{(j)},
\end{eqnarray}
where $E_\mathrm{e}$ is the energy of Rydberg state and $U(r_{i,j})$ describes the interaction between atoms $i$ and $j$ separated by $r_{i,j}$.
Without loss of generality we have set the ground-state energy $E_\mathrm{g} = 0$. By using a relation $P_\mathrm{ee}^{(j)} = \langle P_\mathrm{ee}^{(j)} \rangle + \delta P_\mathrm{ee}^{(j)}$ where $\langle P_\mathrm{ee}^{(j)} \rangle$ and $\delta P_\mathrm{ee}^{(j)}$ are the mean value and fluctuation of $P_\mathrm{ee}^{(j)}$, respectively, we obtain the mean-field approximation from
\begin{eqnarray}
\label{eq:mean2}
P_\mathrm{ee}^{(i)}P_\mathrm{ee}^{(j)} &\approx & \langle P_\mathrm{ee}^{(i)} \rangle \langle P_\mathrm{ee}^{(j)} \rangle + \langle P_\mathrm{ee}^{(i)} \rangle  \delta P_\mathrm{ee}^{(j)} + \langle P_\mathrm{ee}^{(j)} \rangle  \delta P_\mathrm{ee}^{(i)} \\\nonumber
&=& \langle P_\mathrm{ee}^{(i)} \rangle P_\mathrm{ee}^{(j)} + \langle P_\mathrm{ee}^{(j)} \rangle P_\mathrm{ee}^{(i)} - \langle P_\mathrm{ee}^{(i)} \rangle \langle P_\mathrm{ee}^{(j)} \rangle,
\end{eqnarray}
which allows to write the Hamiltonian in the following form
\begin{eqnarray}
\label{eq:mean3}
H &=& \sum_{j=1}^{N} H_{j} \\\nonumber
&=& \sum_{j=1}^{N} \hbar \left( \omega + \left( \langle P_\mathrm{ee}\rangle - \frac{\langle P_\mathrm{ee}\rangle^{2}}{2}    \right)\Delta\omega_{j} \right) P_\mathrm{ee}^{(j)} - \hbar\left(\frac{\langle P_\mathrm{ee}\rangle^{2}}{2} \Delta\omega_{j} \right)P_\mathrm{gg}^{(j)},
\end{eqnarray}
where $\omega=E_\mathrm{e}/\hbar$ is the atomic-resonance frequency and $\langle P_\mathrm{ee}^{(j)} \rangle$ is assumed to be common to all the atoms and set to $\langle P_\mathrm{ee} \rangle$.
Each atom can be considered separately, and the interactions enter as shifts of the energy levels with $\hbar\left( \langle P_\mathrm{ee}\rangle - \frac{\langle P_\mathrm{ee}\rangle^{2}}{2}\right)\Delta\omega_{j}$ and $\hbar\left(\frac{\langle P_\mathrm{ee}\rangle^{2}}{2} \Delta\omega_{j} \right)$ for the Rydberg and ground states, respectively. Here $\Delta\omega_{j} = \sum_{j \neq i} U(r_{i,j})/\hbar$ is the sum over all interactions with atom `$j$'.

In this model, therefore, Rydberg interactions modify only the period of the Ramsey oscillation of each atom as follows
\begin{eqnarray}
\label{eq:mean4a}
P_{\mathrm{mf};j}(\tau) = 2p_\mathrm{g}p_\mathrm{e}\left( 1+ \cos((\omega + p_\mathrm{e}\Delta\omega_{j})\tau + \phi) \right),
\end{eqnarray}
where $\tau$ is the pump-probe delay, $p_\mathrm{g}$ and $p_\mathrm{e}$ are the ground- and Rydberg-state populations, respectively, $\langle P_\mathrm{ee} \rangle=p_\mathrm{e}$, and $\phi$ is the phase offset arising from the AC-Stark shifts during the pulse excitation. Oscillations with slightly different periods are then averaged over the atom distribution.
The interferogram obtained in the Ramsey measurement is thus given by
\begin{eqnarray}
\label{eq:mean4}
P_{\mathrm{mf}}(\tau) = \frac{1}{N} \sum_{j=1}^{N} P_{\mathrm{mf};j}(\tau) = 2p_\mathrm{g}p_\mathrm{e}\left( 1+ \frac{1}{N}\sum_{j=1}^{N}\cos((\omega + p_\mathrm{e}\Delta\omega_{j})\tau + \phi) \right).
\end{eqnarray}
This averaging yields a contrast decay of the Ramsey oscillation.
By employing the Monte Carlo method to model a realistic distribution of the $N$ atoms as given by Supplementary Eq.~(\ref{eq:nni4}) we acquire the results of Supplementary Eq.~(\ref{eq:mean4}). \\

\noindent
{\bf {\large Supplementary Note 8: Outline of the least-squares fitting}}

The least-squares fitting for Fig.~3b in the main text and Supplementary Figs.~\ref{Suppfig04}b and \ref{Suppfig06}b was performed so that the residual between the Ramsey contrasts measured for $\sim$3.3\,\% population and their simulations was minimum, with reasonable steps of fitting parameters, for 500 atoms, whose configuration was generated by the Monte-Carlo simulation. For each of these 500 atoms, the mean-field energy shift is calculated by considering the interactions with $\sim 6 \times 10^{5}$ atoms, and these shifts are averaged over the 500 atoms. The fitting parameters such as the $C_6$ coefficient obtained in that least-squares fitting were used to calculate the simulated curves shown in Fig.~3 and Supplementary Figs.~\ref{Suppfig04} and \ref{Suppfig06} with an averaging over 20,000 atoms from the ensemble.

The least-squares fitting for Fig.~6b in the main text and Supplementary Figs.~\ref{Suppfig05}b and \ref{Suppfig07}b was also performed so that the residual between the Ramsey contrasts measured for $\sim$ 3.3\,\% population and their simulations with Eq~(4) in the main text was minimum with reasonable steps of fitting parameters. The fitting parameters such as the $C_6$ coefficient obtained in that least-squares fitting were used to calculate the simulated curves shown in Fig.~6 and Supplementary Figs.~\ref{Suppfig05} and \ref{Suppfig07}.\\

\noindent
{\bf {\large Supplementary Note 9: Effects of the atom-density distribution on the phase-shifts}}

We present detailed analyses of the effects of the atom-density distribution on the phase-shifts. Supplementary Figure~\ref{Suppfig16}a shows the exact and mean-field calculations of the phase-shift for the 42D Rydberg state with the Gaussian distribution of the atom density. The Gaussian distribution has been taken from the Ramsey experiment shown in the main text.
Both in the exact and mean-field calculations, the phase-shift is saturated as the Rydberg population $p_\mathrm{e}$ increases, as shown in Supplementary Fig.~\ref{Suppfig16}a. As the Rydberg population is increased, the initial slope of the phase-shift becomes larger due to stronger interactions, so the phase-shift converges earlier to a value around -$40$ degrees, and the disagreement between the exact and mean-field results becomes smaller.

This saturation is due to the Gaussian atom-density distribution of our experimental setup as shown in Supplementary Fig.~\ref{Suppfig16}b, in which the phase-shifts calculated with two different density distributions (Gaussian and homogeneous) and two different Rydberg populations are compared within the mean-field model. The Gaussian distribution is the same as in Supplementary Fig.~\ref{Suppfig16}a, whereas the homogeneous density has been set to the average density of the Gaussian distribution.
It is seen that the Gaussian distribution gives the phase-shift saturated more rapidly for the higher Rydberg population.
With the homogeneous distribution, however, the phase-shifts are not saturated within this timescale $1000$\,ps. In contrast to the homogeneous distribution, the rapid decrease of the atom density in the Gaussian tails results in the saturation of the phase-shift. This is because the contribution to the phase-shift from atoms distant from the center of the Gaussian distribution is suppressed, and therefore the phase-shift does not grow afterwards.
The stronger interactions for $p_\mathrm{e} \sim 3.3\,\%$ yield the phase-shift saturated within our measurement time $500$\,ps. This saturation is reached both by the exact and mean-field calculations for $p_\mathrm{e} \sim3.3\,\%$ as seen in Supplementary Fig.~\ref{Suppfig16}a, unlike the case with $p_\mathrm{e} \sim 1.2\,\%$ and therefore weaker interactions, giving the closer agreement between the mean-field and exact results as well as the experimental ones for $p_\mathrm{e} \sim 3.3\,\%$ than for $p_\mathrm{e} \sim 1.2\,\%$, as seen in Fig.~3d in the main text.\\

\noindent
{\bf {\large Supplementary Note 10: Numerical simulations with alternative forms of interactions}}\\
{\bf {\large Supplementary Note 10-1: Effective treatment of a two-atom interaction}}

At short interatomic distances a dipole-dipole interaction in non-diagonal terms of the Hamiltonian couples the initial Rydberg states of a pair of atoms described by $|\cdots \mathrm{e} \cdots \mathrm{e} \cdots\rangle$ with other Rydberg states $|\cdots \mathrm{e}' \cdots \mathrm{e}'' \cdots\rangle$. Such couplings induce hybridization among multiple Rydberg states, leading to congested potential structures.
To handle this intractable problem in the present study, we have considered a single effective potential for the two-atom interaction that represents the congested potential structure. Thereby, the effective interaction enters only the diagonal components of the Hamiltonian.\\

\noindent
{\bf {\large Supplementary Note 10-2: Numerical simulation with an anisotropic van der Waals interaction}}

In the main text we have considered an isotropic van der Waals interaction given by $U(r) = -C_{6}/r^{6}$. Here we introduce an anisotropy into the van der Waals interaction by
\begin{eqnarray}
\label{eq:Aniso1}
U(r,\theta) = -\frac{C_{6}(1-3\cos^{2}(\theta))^{2}}{r^{6}},
\end{eqnarray} 
where $\theta$ is the angle between the z-axis (Fig.~1a) and a line connecting two atoms interacting with each other. The comparisons between our experimental observations and the simulations with the anisotropic potential $U(r,\theta)$ above are presented in Supplementary Fig.~\ref{Suppfig04} for the mean-field model and in Supplementary Fig.~\ref{Suppfig05} for the theory model beyond mean-field with the analytical continuum-approximation, respectively. It is seen in Supplementary Fig.~\ref{Suppfig04}c that the mean-field simulation fails to reproduce the observed phase-shift again, whereas the theory-model simulations agree with the measured Ramsey contrasts and phase-shifts for both of the Rydberg populations $p_\mathrm{e} \sim$1.2$\,\%$ to 3.3$\,\%$. The adjustment parameter employed in these simulations with the anisotropic potential is $C_6=63$\,GHz\,$\mu$m$^6$ and is comparable to $C_6=34$\,GHz\,$\mu$m$^6$ employed in the simulations without the potential anisotropy.\\

\noindent
{\bf {\large Supplementary Note 10-3: Numerical simulation with a hybrid form of a dipole-dipole and a van der Waals interaction and with a pure dipole-dipole interaction}}

A hybrid form of a dipole-dipole and a van der Waals (DD-vdW) interaction is defined by
\begin{eqnarray}
\label{eq:Eff1}
U(r) = \left\{ \begin{array}{cc} -\frac{C_{3}}{r^{3}} & r \leq r_\mathrm{c} \\  -\frac{C_{6}}{r^{6}} & r > r_\mathrm{c} \end{array}, \right.
\end{eqnarray}
where $r_\mathrm{c}$ is the crossover radius defined by $r_\mathrm{c}^{3} = C_{6}/C_{3}$. We have performed numerical simulations of the Ramsey contrast and phase-shift using this DD-vdW interaction with $C_{3}$ and $r_\mathrm{c}$ being fitting parameters.

We have also performed the simulations using a dipole-dipole (DD) interaction without anisotropies. The results of those simulations are presented in Supplementary Fig.~\ref{Suppfig06} for the mean-field model and in Supplementary Fig.~\ref{Suppfig07} for the theory model with the continuum approximation, respectively. The corresponding results with the pure van der Waals~(vdW) interaction, which have been shown in Figs.~3 and 6 of the main text, are presented again in Supplementary Figs.~\ref{Suppfig06} and \ref{Suppfig07} to be compared with the DD-vdW and DD results.
It is seen in Supplementary Fig.~\ref{Suppfig06}c that the mean-field simulation fails to reproduce the observed phase-shift again for both the DD-vdW and DD interactions. It is seen in Supplementary Fig.~\ref{Suppfig07}, on the other hand, that the theory model simulations beyond mean-field with the DD-vdW interaction show good agreements simultaneously with the Ramsey contrasts and phase-shifts, whereas similar agreements have not been found with the DD interaction.
Supplementary Figure~\ref{Suppfig08} shows the two-atom potentials that we have obtained by diagonalizing the $^{87}$Rb Hamiltonian for two atoms with the DD interaction.
It is seen from this figure that there is a mixing of states correlating to the $42$D$_{5/2}$+$42$D$_{5/2}$ and $44$P$_{3/2}$+$40$F$_{7/2}$ asymptotes due to DD coupling at interatomic distances shorter than $\sim2\,\mu$m. This distance is longer than the shortest interatomic distance accessible by our ps pulse excitation, indicating the validity of  the DD-vdW interaction. The DD-vdW interaction, however, needs two fitting parameters $C_{3}$ and $r_\mathrm{c}$, whereas the pure vdW interaction needs only one fitting parameter $C_6$. We have, therefore, employed the pure vdW interaction in the main text to suppress the ambiguity of the fitting.

Supplementary Figure~\ref{Suppfig09} shows the simulated Ramsey contrast and phase-shift at $\tau=$ 500\,ps as functions of the cutoff radius $r_0$ (the lower abscissa), at which we truncate the integration in Eq.~(6) in the main text, and of an average number of interacting atoms within the volume $V=\frac{4\pi}{3}(r_0^3-r_{\mathrm{B}}^3)$ (the higher abscissa). They are simulated by the theory model with the continuum approximation for the vdW, DD, and DD-vdW interactions without anisotropies. The population of the $42$D$_{5/2}$ is set to $\sim$3.3$\%$ in these simulations.
The converged values agree well with the measured ones indicated by the dark-grey solid lines, each of which is the average over eight points around $\tau=500$\,ps in Figs.~6b or d in the main text. The light-grey shaded area represents one standard deviation of the average over those eight measured values. On the other hand, the Ramsey contrast and phase-shift simulated with the DD interaction neither converge nor agree with the experimental observations. The black solid line in Supplementary Fig.~\ref{Suppfig09}a shows the Ramsey contrast $|g(\tau)|$ given by Eq.~(7) in the main text with $\gamma(\tau) = 0$, which gives the situation in which the Ramsey oscillations modulated by the Rydberg-Rydberg interactions are dephased completely. This black solid line thus represents the upper limit of the contrast decay and accordingly the lower limit of the number of atoms $\sim 32$ to reproduce the Ramsey contrast $\sim 0.45$ measured at $\tau = 500$\,ps, irrespective of the potential curves.\\



\noindent
{\bf {\Large Supplementary References}}
\noindent
\begin{table}[h]
\begin{tabular}{lp{17cm}}
$[1]$ & Tanner, P. J., Han, J., Shuman, E. S. \& Gallagher, T. F.
Many-body ionization in a frozen Rydberg gas.
{\it Phys. Rev. Lett.} {\bf 100}, 043002 (2008).\\
$[2]$ & Gallagher, T. F.
{\it Rydberg Atoms.}
(Cambridge Univ. Press, 2005).\\
$[3]$ & Nipper, J. {\it et al.}
Atomic pair-state interferometer: controlling and measuring an interaction-induced phase shift in Rydberg-atom pairs.
{\it Phys. Rev. X} {\bf 2}, 031011 (2012).\\
$[4]$ & Zhou, T., Li, S. \& Jones, R. R.
Rydberg-wave-packet evolution in a frozen gas of dipole-dipole-coupled atoms.
{\it Phys. Rev. A} {\bf 89}, 063413 (2014).
\end{tabular}
\end{table}

\end{document}